\documentclass[preprint,floats,aps,epsfig,nofootinbib,amssymb,showpacs]{revtex4}
\usepackage{color}
\usepackage{graphicx}
\usepackage{tabularx}
\usepackage{amsmath}
\usepackage{amssymb}
\usepackage{datetime}
\usepackage[left]{lineno}
\usepackage{txfonts}
\usepackage{dcolumn}
\usepackage{bm}

\begin{document}

\title{Exploring the singlet scalar dark matter from direct detections and neutrino signals via its annihilation in the Sun}

\author{Wan-Lei Guo}
\email[Email: ]{guowl@itp.ac.cn}

\author{Yue-Liang Wu}
\email[Email: ]{ylwu@itp.ac.cn}

\affiliation{ Kavli Institute for Theoretical Physics China, \\
Key Laboratory of Frontiers in Theoretical Physics, \\
Institute of Theoretical Physics, Chinese Academy of Science,
Beijing 100190, China}

\begin{abstract}

We explore the singlet scalar dark matter (DM) from  direct
detections and high energy neutrino signals generated by the solar
DM annihilation. Two singlet scalar DM models are discussed, one is
the real singlet scalar DM model  as the simple extension of the
standard model (SSDM-SM) with a discrete $Z_2$ symmetry, and another
is the complex singlet scalar DM model  as the simple extension of
the left-right symmetric two Higgs bidoublet model (SSDM-2HBDM) with
$P$ and $CP$ symmetries. To derive the Sun capture rate, we consider
the uncertainties in the hadronic matrix elements and calculate the
spin-independent DM-nucleon elastic scattering cross section. We
find that the predicted neutrino induced upgoing muon fluxes in the
region $3.7 \, {\rm GeV} \leq m_D \leq 4.2$ GeV slightly exceed the
Super-Kamiokande limit in the SSDM-SM. However, this exceeded region
can be excluded by the current DM direct detection experiments. For
the SSDM-2HBDM, one may adjust the Yukawa couplings to avoid the
direct detection limits and enhance the predicted muon fluxes. For
the allowed parameter space of the SSDM-SM and SSDM-2HBDM, the
produced muon fluxes in the Super-Kamiokande and muon event rates in
the IceCube are less than the experiment upper bound and atmosphere
background, respectively.

\end{abstract}

\pacs{95.35.+d, 12.60.-i, 95.55.Vj, 13.15.+g}

\maketitle

\section{Introduction}

The existence of dark matter (DM) is by now well confirmed
\cite{DM1,DM2}. The recent cosmological observations have helped to
establish the concordance cosmological model where the present
Universe consists of about 73\% dark energy, 23\% dark matter and
4\% atoms \cite{WMAP7}. Understanding the nature of dark matter is
one of the most challenging problems in particle physics and
cosmology. Currently, many DM search experiments are under way.
These experiments can be classified as the direct DM searches and
the indirect DM searches. The direct DM detection experiments may
observe the elastic scattering of DM particles with nuclei. The
indirect DM searches are designed to detect the DM annihilation
productions, which include neutrinos, gamma rays, electrons,
positrons, protons and antiprotons. In addition, the collider DM
searches at CERN LHC are complementary to the direct and indirect DM
detection experiments.

The indirect DM searches are usually independent of the direct DM
searches. Namely, one can calculate the DM annihilation signals when
the thermal-average of the annihilation cross section times the
relative velocity $\langle \sigma v \rangle$ and the DM annihilation
productions are known. It is worthwhile to stress that the DM
annihilation signals from the Sun (or Earth) depend on both the
direct DM detection and the indirect DM detection. When the DM
particles elastically scatter with nuclei in the Sun, they may lose
most of their energy and are trapped by the Sun \cite{DM1}. The
solar DM capture rate is related to the DM-nucleon elastic
scattering cross section. These trapped DM particles will be
accumulated in the core of the Sun due to repeated scatters and the
gravity potential. Therefore the Sun is a very interesting place for
us to search the DM annihilation signals
\cite{Barger:2007xf,Liu:2008kz,Hooper:2008cf,Erkoca:2009by,Ellis:2009ka,SUN}.
The DM annihilation rate in the Sun depends on  $\langle \sigma v
\rangle$ and the solar DM distribution. If the DM annihilation rate
reaches equilibrium with the DM capture rate, the solar DM
annihilation rate only depend on the DM-nucleon elastic scattering
cross section. Due to the interactions of the DM annihilation
products in the Sun, only the neutrino can escape from the Sun and
reach the Earth. These high energy neutrinos interact with the Earth
rock or ice to produce upgoing muons which may be detected by the
water Cherenkov detector Super-Kamiokande (SK) \cite{SK} and the
neutrino telescope IceCube \cite{IceCube}.

In this paper, we explore the singlet scalar dark matter from direct
detections and  high energy neutrino signals via the solar DM
annihilation in two singlet scalar DM models. One is the real
singlet scalar DM model as the simple extension of the standard
model (SSDM-SM) \cite{SingletDM,Pospelov,
Gonderinger:2009jp,DAMACoGeNT, Guo:2010hq} and another is the
complex singlet scalar DM model as a simple extension of the
left-right symmetric two Higgs bidoublet model (SSDM-2HBDM)
\cite{Wu:2007kt, Guo:2010vy, Guo:2008si}. In the SSDM-SM,  a real
singlet scalar $S$ with a $Z_2$ symmetry is introduced to extend the
standard model. Although this model is very simple, it is
phenomenologically interesting \cite{SingletDM,Pospelov,
Gonderinger:2009jp,DAMACoGeNT, Guo:2010hq}. In the SSDM-2HBDM, the
imaginary part $S_D$ of a complex singlet scalar field $S =
(S_\sigma + i S_D)/ \sqrt{2}$ with $P$ and $CP$ symmetries can be
the DM candidate \cite{Guo:2008si}. The stability of $S_D$ is
ensured by the fundamental symmetries $P$ and $CP$ of quantum field
theory. In Refs. \cite{Guo:2010hq} and \cite{Guo:2008si}, we have
calculated the spin-independent DM elastic scattering cross section
on a nucleon. In fact, one should consider the uncertainties in the
DM direct detection induced by the uncertainties in the hadronic
matrix elements. Here we consider these uncertainties and
recalculate the spin-independent DM-nucleon elastic scattering cross
section. Then we calculate the neutrino fluxes from the singlet
scalar DM annihilation in the Sun and the neutrino induced upgoing
muon fluxes in the Super-Kamiokande and IceCube. This paper is
organized as follows: In Sec. II, we outline the main features of
the SSDM-SM and SSDM-2HBDM, and give the DM-nucleon elastic
scattering cross section. In Sec. III, we numerically calculate the
differential neutrino energy spectrum generated by per DM pair
annihilation, the DM annihilation rate in the Sun and the neutrino
induced upgoing muon fluxes.  Some discussions and conclusions are
given in Sec. IV.


\section{Constraint on singlet scalar dark matter from direct detections}

\subsection{The real singlet scalar dark matter model as an extension of the SM}
\begin{figure}[htb]
\begin{center}
\includegraphics[scale=0.4]{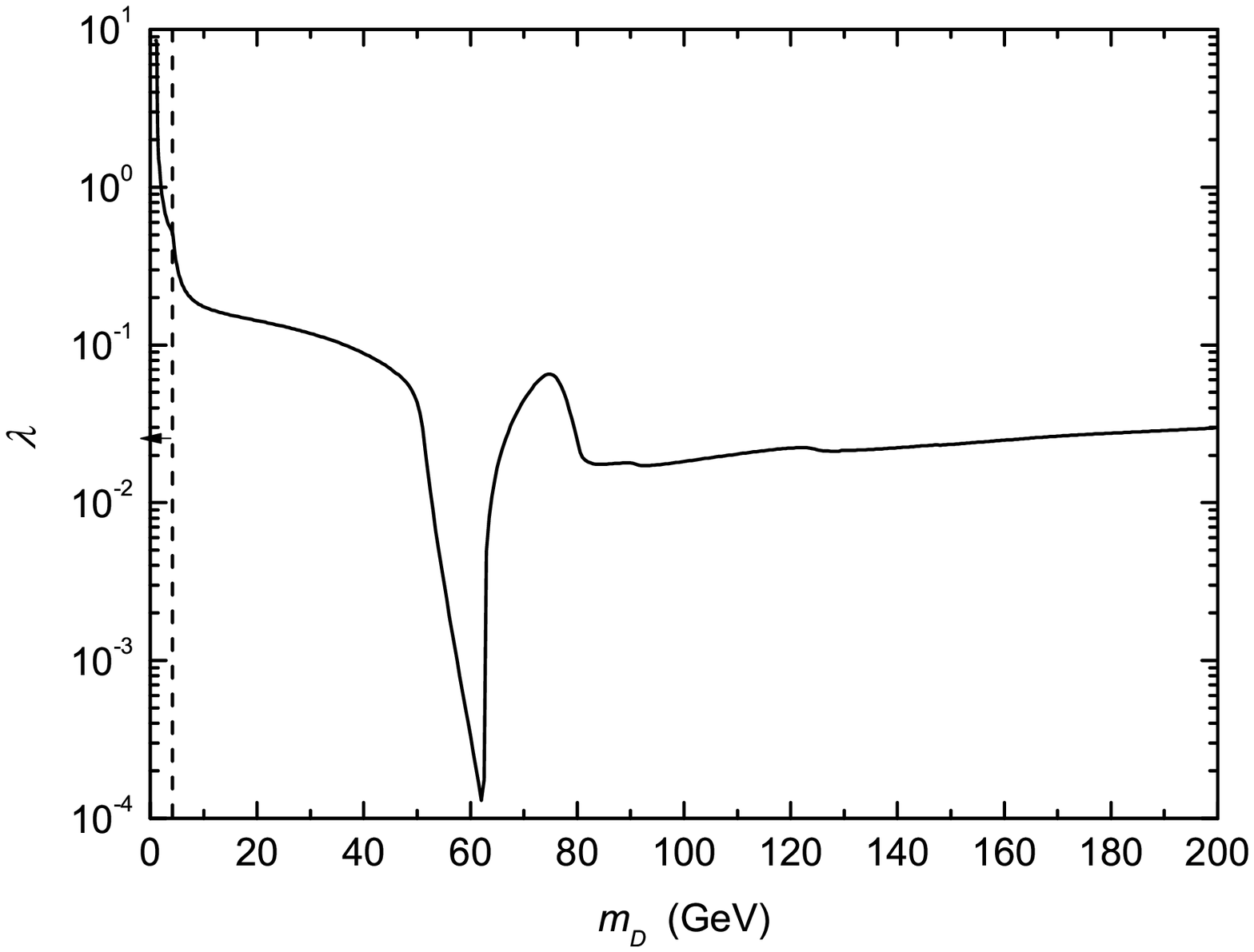}
\includegraphics[scale=0.4]{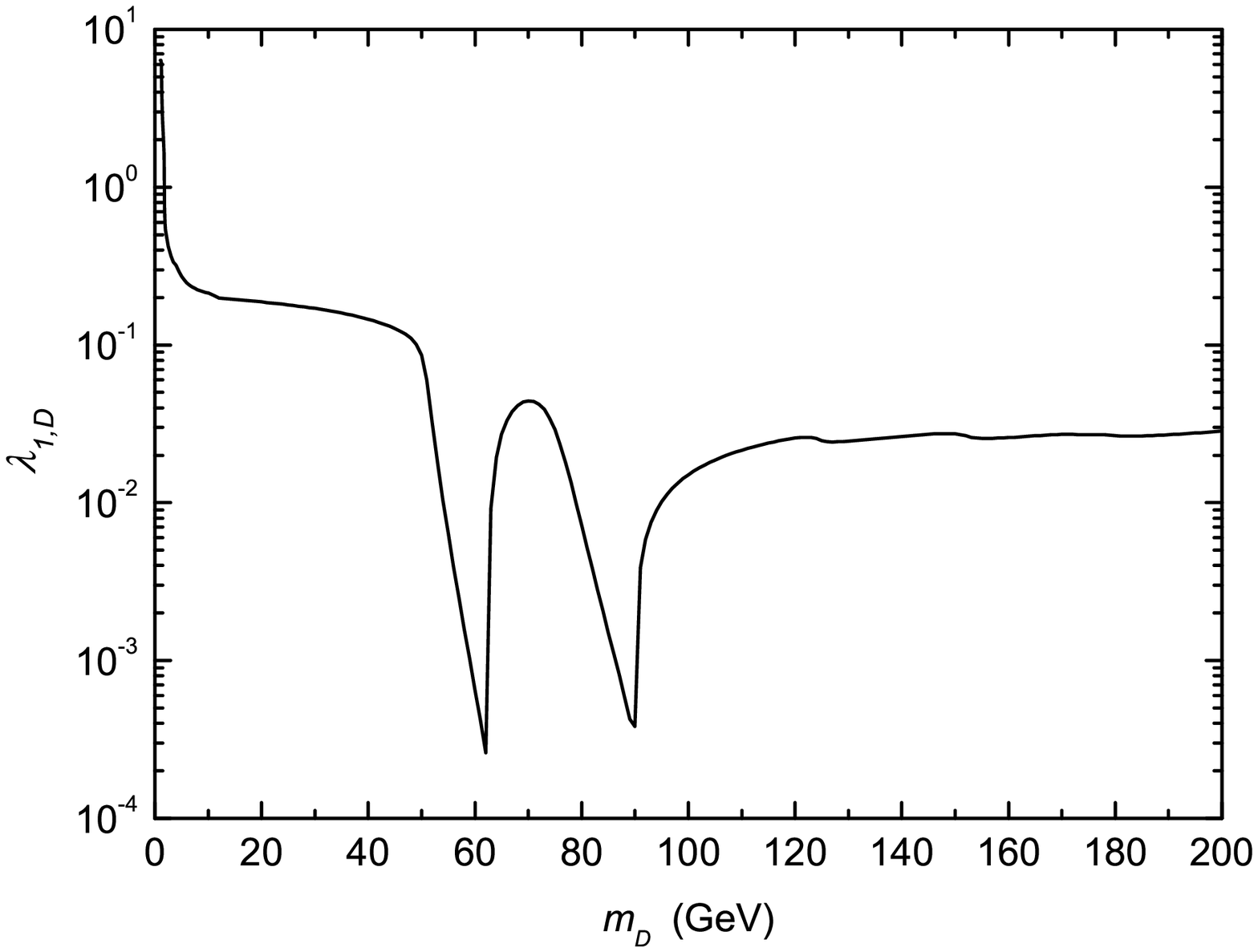}
\end{center}
\caption{ The predicted couplings $\lambda$ in the SSDM-SM (left
panel) and  $\lambda_{1,D}$  in the SSDM-2HBDM (right panel) as a
function of the DM mass $m_D$ from the observed DM abundance
$\Omega_{\rm DM} h^2 = 0.1123\pm0.0035$ \cite{WMAP7}. The vertical
dashed line with arrowhead in the left panel shows the excluded
region from the potential's global minimum, perturbativity and DM
relic density. } \label{lambda}
\end{figure}

In the SSDM-SM, the Lagrangian reads
\begin{eqnarray}
\mathcal{L} = \mathcal{L}_{\rm SM} + \frac{1}{2}\partial_\mu S
\partial^\mu S - \frac{m_0^2}{2} S^2 - \frac{\lambda_S}{4} S^4 - \lambda S^2 H^\dag H \; ,
\end{eqnarray}
where $H$ is the SM Higgs doublet. The linear and cubic terms of the
scalar $S$ are forbidden by the $Z_2$ symmetry $S \rightarrow -S$.
Then $S$ has a vanishing vacuum expectation value (VEV) $\langle S
\rangle =0$ which ensures the DM candidate $S$ stable. $\lambda_S$
describes the DM self-interaction strength which is independent of
the DM annihilation.  After the spontaneous symmetry breaking (SSB),
one can obtain the DM mass $m_D^2 = m_0^2 + \lambda \; v_{\rm EW}^2$
with $v_{\rm EW}= 246$ GeV. The SSDM-SM is very simple and has only
three free parameters: the DM mass $m_D$, the Higgs mass $m_h$ and
the coupling $\lambda$. In terms of the observed DM abundance
$\Omega_{\rm DM} h^2 = 0.1123 \pm 0.0035$ \cite{WMAP7}, one can
calculate the coupling $\lambda$ for the given $m_D$ and $m_h$. Here
we take $m_h = 125$ GeV \cite{LHC} and $1 \;{\rm GeV} \leq m_D \leq
200$ GeV for illustration. As shown in Fig. \ref{lambda}, the
observed DM abundance requires $\lambda \sim {\cal O} (10^{-4} -
1)$. It is well known that the annihilation cross section $\sigma$
will become larger for the same coupling when the annihilation
process nears a resonance. This feature implies that there is a very
small coupling when $0.8 \; m_h \lesssim 2 m_D < m_h$ as shown in
Fig. \ref{lambda}. This region is named as the resonance region in
the following parts of this paper.

\begin{figure}[htb]
\begin{center}
\includegraphics[scale=0.4]{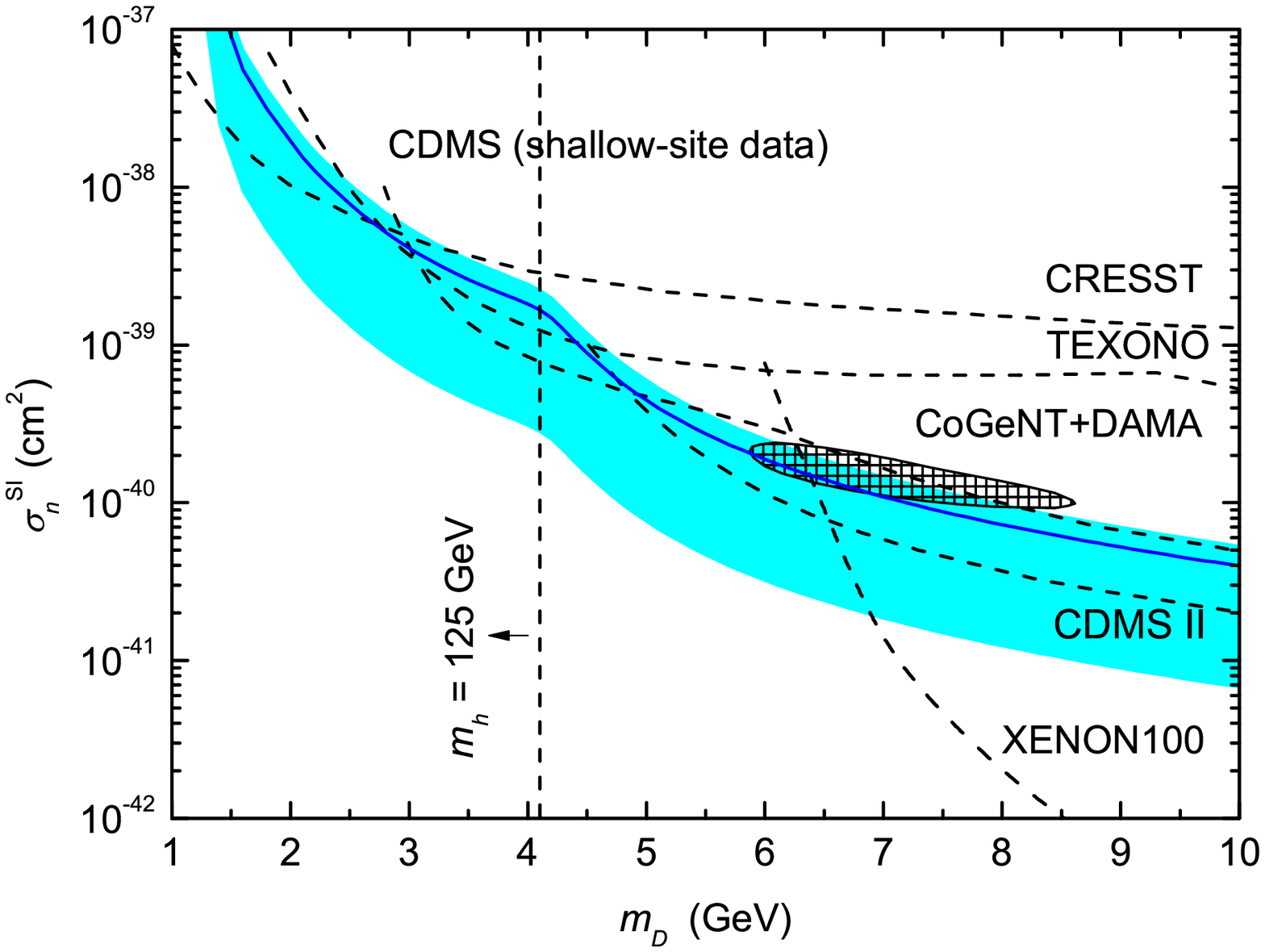}
\includegraphics[scale=0.4]{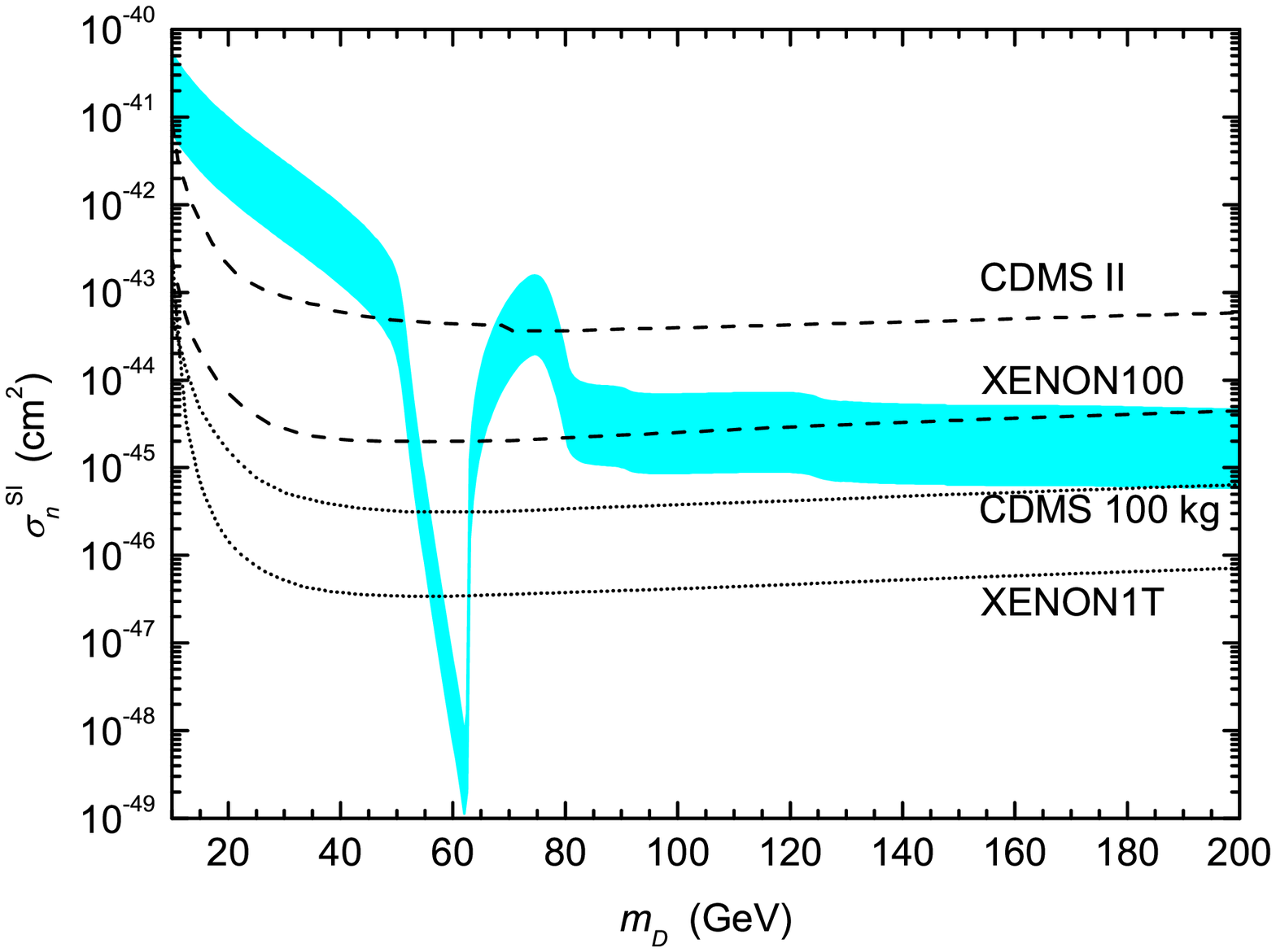}
\includegraphics[scale=0.4]{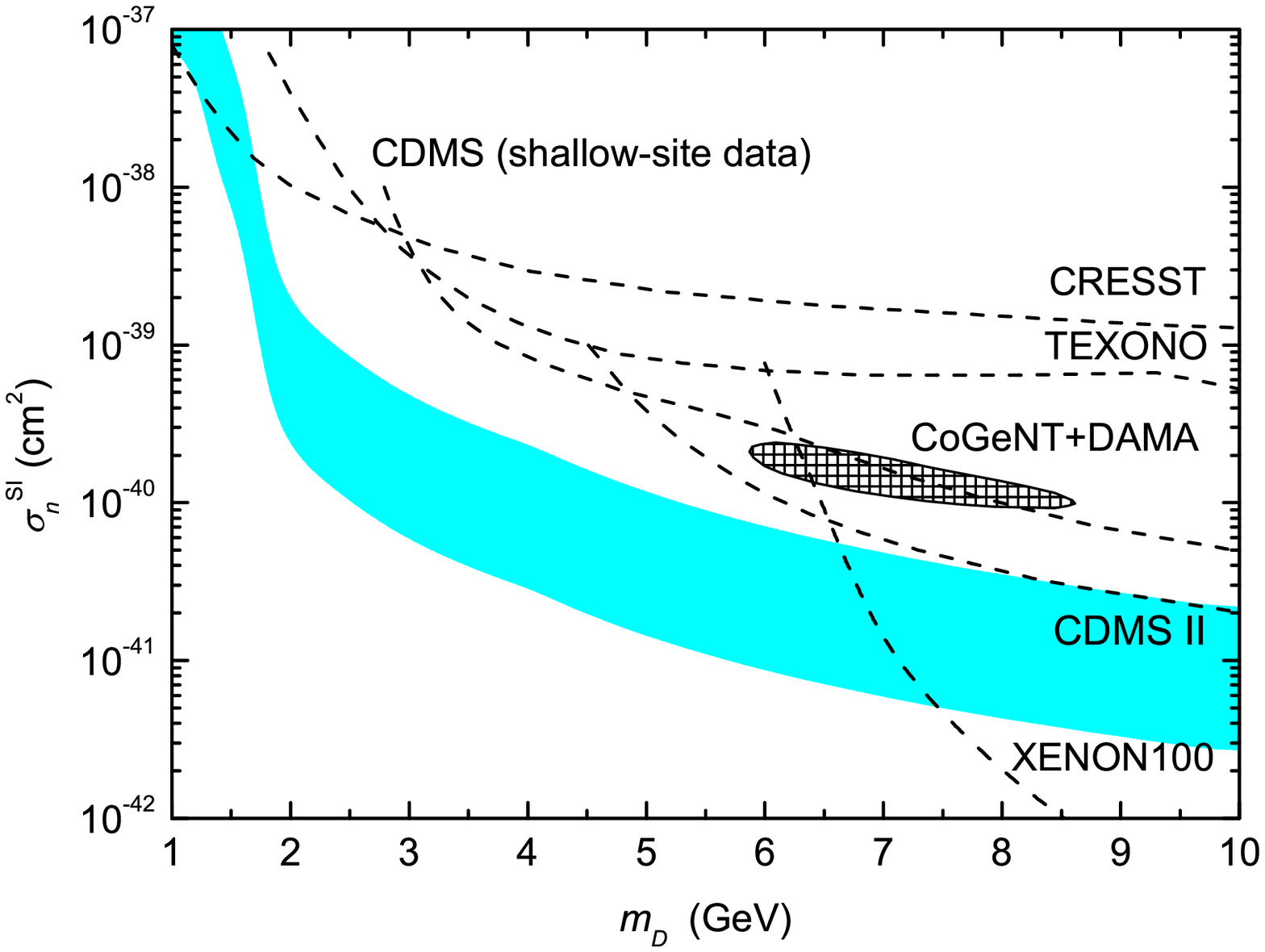}
\includegraphics[scale=0.4]{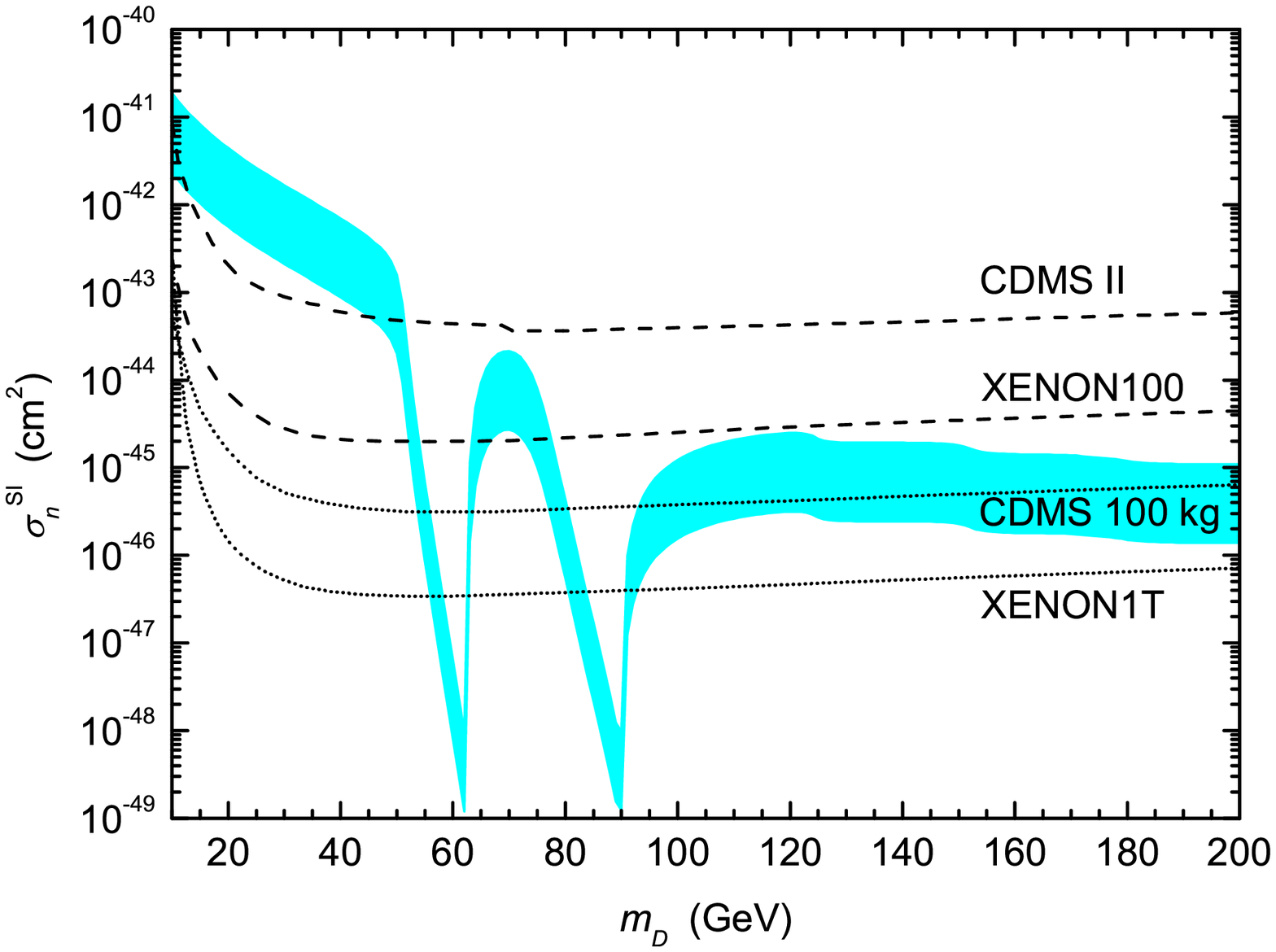}
\end{center}
\caption{ The predicted DM-nucleon elastic scattering cross section
$\sigma_{n}^{\rm SI}$ in the SSDM-SM (top row) and SSDM-2HBDM
(bottom row) for $1 \;{\rm GeV} \leq m_D \leq 200$ GeV. The dashed
lines indicate the current experimental upper bounds. The short
dotted lines in the right panels denote the future experimental
upper bounds from the CDMS 100 kg \cite{CDMS100} and XENON1T
\cite{XENON1T}. The blue solid line in the top left panel describes
the $f=0.63$ case. The black hatched region corresponds to a
combination of the DAMA and CoGeNT \cite{Hooper:2010uy}.  The
vertical dashed line with arrowhead in the top left panel shows the
excluded region from the potential's global minimum, perturbativity
and DM relic density. } \label{Direct}
\end{figure}

Using the predicted $\lambda$ from the observed DM abundance, one
can calculate the spin-independent DM-nucleon elastic scattering
cross section \cite{miss}
\begin{eqnarray}
\sigma_{n}^{\rm SI} \approx \frac{\lambda^2}{\pi} f^2
\frac{m_n^2}{m_h^4 m_D^2}\left (\frac{m_D \; m_n}{m_D+ m_n }
\right)^2 \; , \label{sigmaSM}
\end{eqnarray}
where $m_n$ is the nucleon mass and $f = (7/9)\sum_{q=u,d,s}
f_{Tq}^p+  2/9$. In terms of the relevant formulas in Ref.
\cite{Ellis:2009ka}, one can calculate the parameters $f_{Tq}^p$ and
obtain $f \approx 0.56 \pm 0.17$. On the other hand, the lattice
results imply $f \approx 0.29 \pm 0.03$ where we take the
strange-quark sigma term $16 \;{\rm MeV} \leq \sigma_s \leq 69$ MeV
\cite{Giedt:2009mr}. Therefore we adopt $0.26 \leq f \leq 0.73$ for
the following analyses. The authors in Ref. \cite{DAMACoGeNT} have
discussed that the light DM particle $S$ can explain the DAMA
\cite{DAMA} and CoGeNT \cite{CoGeNT} experiments. Here we consider
the latest experiment limits and recalculate the spin-independent
DM-nucleon elastic scattering cross section $\sigma_{n}^{\rm SI}$
with  $0.26 \leq f \leq 0.73$. Notice that $\sigma_{n}^{\rm SI}$ is
not sensitive to the Higgs mass in the low DM mass range. As shown
in Fig. \ref{Direct} (top left panel), the predicted
$\sigma_{n}^{\rm SI}$ in the region $6 \;{\rm GeV} \lesssim m_D
\lesssim 8$ GeV and $f \gtrsim 0.60$ well fit the common region of
the DAMA and CoGeNT \cite{Hooper:2010uy}. However, the recent CDMS
II \cite{CDMSII} disfavors the CoGeNT+DAMA region. We find that the
CDMS II \cite{CDMSII}, CDMS (shallow-site data) \cite{CDMS}, CRESST
\cite{CRESST} and TEXONO \cite{TEXONO} can exclude the $f \gtrsim
0.63$ region for $1 \;{\rm GeV} \leq m_D \leq 10$ GeV. The $f=0.63$
case has been shown as the blue solid line in the top left panel of
Fig. \ref{Direct}. The latest XENON100 \cite{XENON100} may exclude
$7 \;{\rm GeV} \lesssim  m_D \lesssim 52$ GeV and a narrow region
$65 \;{\rm GeV} \lesssim  m_D \lesssim 80$ GeV even if we take $f =
0.26$ as shown in Fig. \ref{Direct}. The future experiments CDMS 100
kg \cite{CDMS100} and XENON1T \cite{XENON1T} can cover most parts of
the allowed parameter space. As the DM mass increases, new DM
annihilation channels will be open which means that the predicted
$\lambda$  from the DM relic density will quickly decrease.
Therefore a kink around the bottom quark mass $m_D \approx m_b =
4.2$ GeV occurs in the top left panel of Fig. \ref{Direct}. When the
DM mass approaches the resonant point $m_D$ = 62.5 GeV for $m_h$
=125 GeV, one can obtain a very large thermally averaged
annihilation cross section $\langle \sigma v \rangle$ for the given
$\lambda$. In order to derive the correct DM relic density, we have
to require that the coupling $\lambda$ is very small. In this case,
a very small $\sigma_{n}^{\rm SI}$ around the resonant point can be
obtained.

The SSDM-SM also suffers other constraints except for the direct
detections, such as the potential's global minimum at $\langle h
\rangle =v_{\rm EW}$ and $\langle S \rangle =0$ requires $|\lambda|
< \sqrt{\lambda_S/2} m_h/ v_{\rm EW} + m_D^2/v_{\rm EW}^2 $
\cite{Pospelov}. Since the perturbativity implies $6 \lambda_S < 4
\pi$, one can derive $|\lambda| < \sqrt{\pi/3} m_h / v_{\rm EW} +
m_D^2/v_{\rm EW}^2$. Then we find the desired DM relic density can
exclude $m_D \lesssim 4.1$ GeV for $m_h = 125$ GeV. The vertical
dashed line with arrowhead in the left panels of Figs. \ref{lambda}
and \ref{Direct} shows the excluded region. In Ref.
\cite{Gonderinger:2009jp}, the authors have given the lower bounds
on $m_D$ for several typical $\lambda_S$ based on the one-loop
vacuum stability and the observed DM relic density.

\subsection{The complex singlet scalar dark matter model as an extension of the 2HBDM}

We begin with a brief review of the 2HBDM described in Ref.
\cite{Wu:2007kt}.  The model is based on the gauge group
$SU(2)_L\otimes SU(2)_R\otimes U(1)_{B-L}$. The left- and
right-handed fermions belong to $SU(2)_L$ and $SU(2)_R$ doublets,
respectively. The Higgs sector  contains two Higgs bidoublets $\phi$
(2,$2^{*}$,0), $\chi$ (2,$2^{*}$,0) and a left(right)-handed Higgs
triplet $\Delta_{L(R)}$ (3(1),1(3),2) with the following flavor
contents
\begin{eqnarray}
\phi  = \left ( \begin{matrix} \phi_1^0 & \phi_2^+ \cr \phi_1^- &
\phi_2^0 \cr  \end{matrix} \right ) , \; \chi  = \left (
\begin{matrix} \chi_1^0 & \chi_2^+ \cr \chi_1^- & \chi_2^0 \cr
\end{matrix} \right ) , \; \Delta_{L,R}  = \left (
\begin{matrix} \delta_{L,R}^+/\sqrt{2} & \delta_{L,R}^{++} \cr
\delta_{L,R}^{0} & -\delta_{L,R}^{+}/\sqrt{2} \cr \end{matrix}
\right ) . \label{Higgscomponent}
\end{eqnarray}
The introduction of Higgs bidoublets $\phi$ and $\chi$ can account
for the electroweak symmetry breaking and overcome the fine-tuning
problem in generating the spontaneous $CP$ violation  in the
left-right symmetric one Higgs bidoublet model. Meanwhile it also
relaxes the severe low energy phenomenological constraints
\cite{Wu:2007kt}. Motivated by the spontaneous $P$ and $CP$
violations, we require $P$ and $CP$ invariance of the Lagrangian,
which strongly restricts the structure of the Higgs potential. The
most general potential containing only the $\phi$ and $\Delta_{L,R}$
fields is given by
\begin{eqnarray}
\mathcal{V}_{\phi\Delta} & = & -\mu_{1}^{2}\rm{Tr}(\phi^{\dagger}\phi)-\mu_{2}^{2}[\rm{Tr}(\tilde{\phi}^{\dagger}\phi)+\rm{Tr}(\tilde{\phi}\phi^{\dagger})]-\mu_{3}^{2}[\rm{Tr}(\Delta_{L}\Delta_{L}^{\dagger})+\rm{Tr}(\Delta_{R}\Delta_{R}^{\dagger})]\nonumber\\
 &  & +\lambda_{1}[\rm{Tr}(\phi^{\dagger}\phi)]^{2}+\lambda_{2}\{[\rm{Tr}(\tilde{\phi}^{\dagger}\phi)]^{2}+[\rm{Tr}(\tilde{\phi}\phi^{\dagger})]^{2}\}+\lambda_{3}[\rm{Tr}(\tilde{\phi}^{\dagger}\phi)\rm{Tr}(\tilde{\phi}\phi^{\dagger})]\nonumber\\
 &  & +\lambda_{4}\{\rm{Tr}(\phi^{\dagger}\phi)[\rm{Tr}(\tilde{\phi}^{\dagger}\phi)+\rm{Tr}(\tilde{\phi}\phi^{\dagger})]\}\nonumber\\
 &  & +\rho_{1}\{[\rm{Tr}(\Delta_{L}\Delta_{L}^{\dagger})]^{2}+[\rm{Tr}(\Delta_{R}\Delta_{R}^{\dagger})]^{2}\}+\rho_{2}[\rm{Tr}(\Delta_{L}\Delta_{L})\rm{Tr}(\Delta_{L}^{\dagger}\Delta_{L}^{\dagger})+\rm{Tr}(\Delta_{R}\Delta_{R})\rm{Tr}(\Delta_{R}^{\dagger}\Delta_{R}^{\dagger})]\nonumber\\
 &  & +\rho_{3}[\rm{Tr}(\Delta_{L}\Delta_{L}^{\dagger})\rm{Tr}(\Delta_{R}\Delta_{R}^{\dagger})]+\rho_{4}[\rm{Tr}(\Delta_{L}\Delta_{L})\rm{Tr}(\Delta_{R}^{\dagger}\Delta_{R}^{\dagger})+\rm{Tr}(\Delta_{L}^{\dagger}\Delta_{L}^{\dagger})\rm{Tr}(\Delta_{R}\Delta_{R})]\nonumber\\
 &  & +\alpha_{1}\rm{Tr}(\phi^{\dagger}\phi)[\rm{Tr}(\Delta_{L}\Delta_{L}^{\dagger})+\rm{Tr}(\Delta_{R}\Delta_{R}^{\dagger})]
          +\alpha_{2}\rm{Tr}[ (\tilde{\phi}^{\dagger}\phi)+(\tilde{\phi}\phi^{\dagger})]\rm{Tr}[(\Delta_{L}\Delta_{L}^{\dagger})+(\Delta_{R}\Delta_{R}^{\dagger})]\nonumber\\
 &  & +\alpha_{3}[\rm{Tr}(\phi\phi^{\dagger}\Delta_{L}\Delta_{L}^{\dagger})+\rm{Tr}(\phi^{\dagger}\phi\Delta_{R}\Delta_{R}^{\dagger})]\nonumber\\
 &  & +\beta_{1}[\rm{Tr}(\phi\Delta_{R}\phi^{\dagger}\Delta_{L}^{\dagger})+\rm{Tr}(\phi^{\dagger}\Delta_{L}\phi\Delta_{R}^{\dagger})]+\beta_{2}[\rm{Tr}(\tilde{\phi}\Delta_{R}\phi^{\dagger}\Delta_{L}^{\dagger})+\rm{Tr}(\tilde{\phi}^{\dagger}\Delta_{L}\phi\Delta_{R}^{\dagger})]\nonumber\\
 &  & +\beta_{3}[\rm{Tr}(\phi\Delta_{R}\tilde{\phi}^{\dagger}\Delta_{L}^{\dagger})+\rm{Tr}(\phi^{\dagger}\Delta_{L}\tilde{\phi}\Delta_{R}^{\dagger})] ,
\label{Vphidelta}
\end{eqnarray}
where the coefficients $\mu_i$, $\lambda_i$, $\rho_i$, $\alpha_i$
and $\beta_i$ in the potential are all real as all the terms are
self-Hermitian. The Higgs potential $\mathcal{V}_{\chi\Delta}$
involving $\chi$ field can be obtained by the replacement
$\chi\leftrightarrow \phi$ in Eq. (\ref{Vphidelta}). The mixing term
$\mathcal{V}_{\chi\phi\Delta}$ can be obtained by replacing one of
$\phi$ by $\chi$ in all the possible ways in Eq. (\ref{Vphidelta}).
After the SSB, the Higgs multiplets obtain  nonzero VEVs
\begin{equation}
\langle\phi_{1,2}^0\rangle=\frac{\kappa_{1,2}}{\sqrt2} \;,
\langle\chi_{1,2}^0\rangle=\frac{w_{1,2}}{\sqrt2} \; \mbox{ and }\;
\langle\delta_{L,R}^0\rangle=\frac{v_{L,R}}{\sqrt2}\;,
\end{equation}
where $\kappa_1$, $\kappa_2$, $w_1$, $w_2$, $v_L$ and $v_R$ are in
general complex, and $ \kappa \equiv \sqrt{|\kappa_1|^2 +
|\kappa_2|^2+|w_1|^2 + |w_2|^2} \approx 246$ GeV represents the
electroweak symmetry breaking scale.   The value of $v_R$ sets the
scale of left-right symmetry breaking which is directly linked to
the right-handed gauge boson masses. $v_R$  is subjected to strong
constraints from the $K$, $B$ meson mixing as well as low energy
electroweak interactions. The kaon mass difference and the indirect
$CP$ violation quantity $\epsilon_K$ set a bound for $v_R$ around
$10$ TeV \cite{Pospelov:1996fq}. In general, the 2HBDM includes
three light neutral Higgs bosons  and a pair of charged light Higgs
particles, whose masses are order of the electroweak energy scale.
For simplicity, we consider $\kappa_2 \sim w_2 \sim 0$. Then one can
derive three light neutral Higgs bosons: $h, H, A$ from $\phi_1^0$
and $\chi_1^0$, and a pair of charged light Higgs particles: $h^\pm$
from $\chi_1^\pm$. For a concrete numerical illustration, we choose
all the masses $m_{H}$, $m_{A}$, $m_{h^\pm}= 180$ GeV and
$m_{h}=125$ GeV.

\begin{table}[htb]
\begin{center}
\begin{tabular}{|c|c|c||c|c|c||c|c|c|}
\hline   &  $P$  &   $CP$  & &\;\; $P$\;\; & \;$CP$\; &  & \;\; $P$\;\; & \;$CP$\; \\
\hline $S$  &  $S$ &   $S^*$  & $S + S^*$  & + & + & $S - S^*$  & + & - \\
\hline $\phi$ \;   &  $\phi^\dagger$ \; & $\phi^*$  & $S S^*$ & + & + &  Tr($\phi^{\dag} \phi$) & + & + \\
\hline $\tilde{\phi}$ \;   &  $\tilde{\phi}^\dagger$ \; &$\tilde{\phi}^*$ & Tr($\phi^{\dag} \tilde{\phi} \pm \tilde{\phi}^{\dag} \phi $)  & $\pm$ & $\pm$ & Tr($\chi^{\dag} \tilde{\chi} \pm \tilde{\chi}^{\dag} \chi $)  & $\pm$ & $\pm$ \\
\hline $\chi$ \;   &  $\chi^\dagger$ \; &$\chi^*$ & Tr($\chi^{\dag} \tilde{\phi} \pm \tilde{\phi}^{\dag} \chi $)  & $\pm$ & $\pm$ & Tr($\phi^{\dag} \chi \pm \chi^{\dag} \phi $)  & $\pm$ & $\pm$ \\
\hline $\Delta_{L(R)}$   & $\Delta_{R(L)}$  & $\Delta_{L(R)}^*$ &
Tr($\Delta_L^{\dag} \Delta_L + \Delta_R^{\dag} \Delta_R$) & + & + &
Tr($\Delta_L^{\dag} \Delta_L - \Delta_R^{\dag} \Delta_R$) &
- & + \\
\hline
\end{tabular}
\end{center}
\vspace{-0.2cm} \caption{The $P$ and $CP$ transformation properties
of the Higgs particles and their gauge-invariant combinations. The
``+" and ``-" denote even and odd, respectively.  } \label{PCP}
\end{table}

In the 2HBDM, the $P$ and $CP$ symmetries have been required to be
exactly conserved before the SSB, thus the discrete symmetries $P$
and $CP$ can be used to stabilize the DM candidate. In the framework
of 2HBDM with a complex singlet scalar $S = (S_\sigma+i S_D)/\sqrt2$
(SSDM-2HBDM), we have considered this possibility in Ref.
\cite{Guo:2008si}. The $P$ and $CP$ transformation properties  of
the Higgs particles and their gauge-invariant combinations  have
been shown in Table \ref{PCP}. It is clear that the odd powers of
$(S-S^*)$ are forbidden by the $P$ and $CP$ symmetries. This hidden
discrete $Z_2$ symmetry on $S_D$ is induced from the original $P$
and $CP$ symmetries. With the help of this hidden $Z_2$ symmetry,
one may derive $\langle S_D \rangle =0$ or $\langle S_D \rangle \neq
0$ for the VEV of $S_D$. Since the $\langle S_D \rangle \neq 0$ case
means that $S_D$  may decay and can not be the DM candidate, we
require that $S$ obtains a real VEV $ \langle S \rangle =
v_\sigma/\sqrt{2}$. Although both $P$ and $CP$ are broken after the
SSB, there is still a residual $Z_2$ symmetry on $S_D$. Therefore
$S_D$ is a stable particle and can be the DM candidate. We have
checked that the $P$ and $CP$ transformation rules for $S$ defined
in Table \ref{PCP} is actually the only possible way for the
implementation of the DM candidate.

For the annihilation cross section of approximately weak strength,
we expect that the DM mass is in the range of a few GeV and a few
hundred GeV. However, the mass $m_D$ of $S_D$ is related to the LR
symmetry breaking scale  $v_R \sim 10$ TeV. To have a possible light
DM mass, we may consider an approximate global $U(1)$ symmetry on
$S$, i.e. $S\to e^{i\delta}S$. Then the $P$ and $CP$ invariant Higgs
potential involving the singlet $S$ is given by
\begin{eqnarray}
\mathcal{V}_{S}& = & - \mu_D^2 SS^* + \lambda_D (SS^*)^2 +
\sum_{i=1}^7 \lambda_{i,D} SS^*  O_i - \frac{ m_D^2}{4} (S -
S^*)^2\;,\label{VS6}
\end{eqnarray}
where $O_1={\rm{Tr}}(\phi^{\dag}\phi)$,
$O_2={\rm{Tr}}(\phi^{\dag}\tilde\phi+\tilde\phi^{\dag}\phi)$,
$O_3={\rm{Tr}}(\chi^{\dag}\chi)$,
$O_4={\rm{Tr}}(\chi^{\dag}\tilde\chi+\tilde\chi^{\dag}\chi)$,
$O_5={\rm{Tr}}(\phi^{\dag}\chi+\chi^{\dag}\phi)$,
$O_6={\rm{Tr}}(\chi^{\dag} \tilde\phi + \tilde\phi^{\dag}\chi)$ and
$O_7={\rm{Tr}}(\Delta_L^{\dag}\Delta_L+\Delta_R^{\dag}\Delta_R)$.
Only the last term explicitly violates $U(1)$ symmetry. After the
SSB, $S$ obtains a real VEV $v_\sigma/\sqrt{2}$. Then one can
straightly derive
\begin{eqnarray}
\mathcal{V}_{S}  =  \frac{\lambda_{D}}{4} [(S_{\sigma}^2 + 2
 v_\sigma S_\sigma  + S_D^2)^2  -  v_\sigma ^4]  + \sum_{i=1}^7 \frac{\lambda_{i,D}}{2} (S_{\sigma}^2 + 2
 v_\sigma S_\sigma  + v_\sigma ^2 + S_D^2) (O_{i} - \langle O_i\rangle)
+ \frac{m_D^2}{2}  S_D^2 \,, \label{VS}
\end{eqnarray}
where we have used the minimization condition $\mu_{D}^2 =
\lambda_{D} v_{\sigma}^2 + \sum_{i} \lambda_{i,D}\langle O_i\rangle$
from the singlet $S_{\sigma}$ to eliminate the  parameter $\mu_D$.
The terms proportional to odd powers of $S_D$ are absent in Eq.
(\ref{VS}) which implies $S_D$ can only be produced by pairs. Notice
that the mass term of $S_D$ should be absent with an exact global
$U(1)$ symmetry.  As discussed in Ref. \cite{Guo:2008si}, the
explicit breaking of this $U(1)$ symmetry can explain the
naturalness of a light DM mass $m_D$, but it does not destroy the
stability of the DM candidate $S_D$. For the VEV of $S_\sigma$, we
require $v_\sigma > v_R \sim 10 \; {\rm TeV} \gg \kappa$ which means
the mixing angles between $S_\sigma$ and other neutral Higgs bosons
in the SSDM-2HBDM are small and the mass of $S_\sigma$ is very
heavy.

\begin{figure}[htb]
\begin{center}
\includegraphics[scale=1.2]{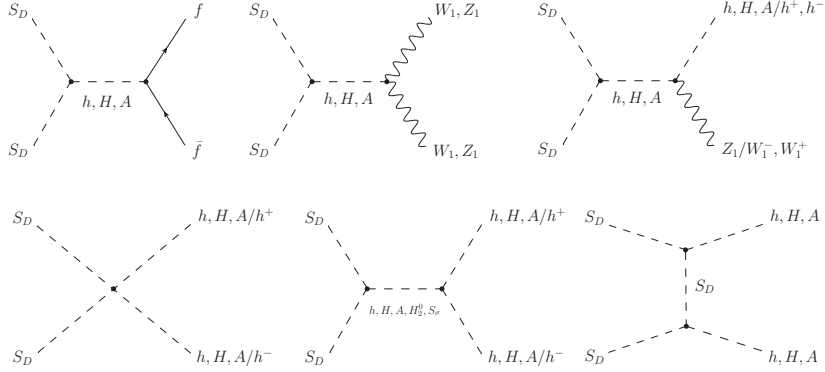}
\end{center}
\caption{ Feynman diagrams for the DM annihilation in the
SSDM-2HBDM. } \label{Feynman2}
\end{figure}

For the DM mass, we take  $1 \;{\rm GeV} \leq m_D \leq 200 $ GeV. In
this case, the possible DM annihilation products are $f \bar f$,
$W_1 W_1 / Z_1 Z_1$, $W_1^\pm h^\mp/ Z_1 (h, H, A)$, $h^+ h^-$ and
any two of the three neutral states $(h,H,A)$ as shown in Fig.
\ref{Feynman2}. Here $W_1$ and $Z_1$ denote the SM gauge bosons. For
cubic and quartic scalar vertexes, we assume they are the same as
those in the one Higgs bidoublet case \cite{Guo:2008si}. Namely, the
vertexes of $S_D S_D (h, H, A)$ and $S_D S_D (h, H, A/h^+) (h, H,
A/h^-)$ are set equal to $- i \lambda_{1, D} v_{\rm EW}$ and $- i
\lambda_{1, D}$, respectively. Similarly, the cubic scalar vertexes
among the light Higgs particles $h$, $H, A$ and $h^{\pm}$ are set
equal to $- i 3 m_{h}^2/v_{\rm EW}$, and the cubic scalar vertexes
between $S_\sigma$ and two light Higgs particles are assumed to be
$- i \lambda_{1,D} v_\sigma$. The vertexes of $f \bar{f} (h, H, A)$
are related with the light Higgs mixing and the Yukawa scale factors
$R_q$. $R_q$ controls the Yukawa couplings and its definition can be
found in Eq. (28) of Ref. \cite{Guo:2008si}. Once the light Higgs
mixing and $R_q$ are fixed, one can predict the coupling
$\lambda_{1,D}$ from the DM relic density.

In the SSDM-2HBDM, the DM-nucleon elastic scattering cross section
is given by \cite{Guo:2008si}
\begin{eqnarray}
\sigma_{n}^{\rm SI} \approx \frac{\lambda_{1,D}^2}{4 \pi} f^2
\frac{m_n^2}{m_D^2}  \left (\frac{m_D \; m_n}{m_D+ m_n } \right)^2
\left ( \frac{f_1}{m_h^2} +\frac{f_3}{m_H^2} +\frac{f_5}{m_A^2}
\right)^2  \; . \label{sigma2HBDM}
\end{eqnarray}
The parameters $f_1, f_3$ and $f_5$ have been given in Ref.
\cite{Guo:2008si} and  are related with the light Higgs mixing and
the Yukawa scale factors $R_q$.  Neglecting possible cancelation due
to the light Higgs mixing in Eq. (\ref{sigma2HBDM}), we find that
$\sigma_{n}^{\rm SI}$ can be enhanced by the large $R_q$ and
approach the current experimental upper bound for the heavy DM mass
\cite{Guo:2008si}. On the other hand, one can also adjust $R_q$ to
avoid the current direct detection limits for the light DM mass. For
illustration, we take the Yukawa scale factors $R_q =1$ for quarks
and $R_l =10$ for charged leptons. Meanwhile, we consider the case
II for the light Higgs mixing \cite{Guo:2008si}. In terms of the
observed DM relic density, we calculate the allowed coupling
$\lambda_{1,D}$ for $1 \;{\rm GeV} \leq m_D \leq 200$ GeV. As shown
in Fig. \ref{lambda} (right panel), the observed DM abundance
requires $\lambda \sim {\cal O} (10^{-4} - 1)$. Then we plot the
predicted $\sigma_{n}^{\rm SI}$ with $0.26 \leq f \leq 0.73$ in Fig.
\ref{Direct} (bottom row).  The latest XENON100 \cite{XENON100} may
exclude $7.5 \;{\rm GeV} \lesssim  m_D \lesssim 52$ GeV and $67
\;{\rm GeV} \lesssim  m_D \lesssim 72$ GeV. It is clear that the
SSDM-2HBDM has smaller $\sigma_{n}^{\rm SI}$ than that in the
SSDM-SM for $1 \;{\rm GeV} \leq m_D \leq 10$ GeV. In the next
section, we shall see another advantage of the SSDM-2HBDM. Namely,
the SSDM-2HBDM can give larger neutrino induced upgoing muon fluxes
than those in the SSDM-SM even if the SSDM-2HBDM has the smaller
$\sigma_{n}^{\rm SI}$. This is because that the large $R_l$ can
significantly change the branching ratios of the dominant DM
annihilation channels which are relevant to the produced neutrino
fluxes.

\section{Neutrino signals from the dark matter annihilation in the Sun}

Based on the DM mass $m_D$ discussed in this paper, two DM particles
may annihilate into fermion pairs, gauge boson pairs and Higgs
pairs. Therefore the differential muon neutrino  energy spectrum at
the surface of the Earth from per DM pair annihilation in the Sun
can be written as
\begin{eqnarray}
\frac{d N_{\nu_{\mu}} }{d E_{\nu_{\mu}} } = \sum_{fs}  B_{fs}
\frac{d N^{fs}_{\nu_{\mu}} }{d E_{\nu_{\mu}} } \;,
\end{eqnarray}
where $fs$ denotes the DM annihilation final state and $B_{fs}$ is
the branching ratio into the final state $fs$. $B_{fs}$ can be
exactly calculated when the couplings $\lambda$ and $\lambda_{1, D}$
are obtained from the DM relic density. $d N^{fs}_{\nu_\mu}/d
E_{\nu_\mu}$ is the energy distribution of neutrinos at the surface
of the Earth produced by the final state $fs$ through hadronization
and decay processes in the core of the Sun. It should be mentioned
that some produced particles, such as $B$ mesons and muons, can lose
a part of energy or the total energy before they decay due to their
interactions in the Sun. In addition, we should consider the
neutrino interactions in the Sun and neutrino oscillations. In this
paper, we use the program package WimpSim \cite{WimpSim} to
calculate $d N^{fs}_{\nu_\mu}/d E_{\nu_\mu}$ with the help of Pythia
\cite{Pythia}, Nusigma \cite{Nusigma} and DarkSUSY \cite{DarkSUSY}.
The Pythia can help us to simulate the hadronization and decay of
the annihilation products and collect the produced neutrinos and
antineutrinos. The Nusigma is a neutrino-nucleon scattering Monte
Carlo package for neutrino interactions on the way out of the Sun.
The density profile of the Sun may affect neutrino oscillations due
to matter effects. For the solar density, the WimpSim uses the
standard solar model BS05(OP) \cite{Bahcall:2004pz} which is coded
into the DarkSUSY. Notice that the WimpSim does not simulate the
Higgs annihilation channel. Since the Higgs decay branching ratios
and the energy distribution of the Higgs decay products can be
exactly calculated in the SSDM-SM, the differential neutrino energy
spectrum from the Higgs annihilation channel can be generated by
those from other annihilation channels. Except for the DM masses and
annihilation channels, the WimpSim only requires inputs of the
neutrino oscillation parameters. Here we consider the lastest Daya
Bay results \cite{An:2012eh} and take \cite{Tortola:2012te}
\begin{eqnarray}
\sin^2 \theta_{12} = 0.32 ,\;\; \sin^2 \theta_{23} = 0.49, & &
\sin^2 \theta_{13} = 0.026, \;\; \delta=0.83 \pi, \nonumber \\
\Delta m_{21}^2 = 7.62 \times 10^{-5} {\rm eV}^2,  & & \Delta
m_{31}^2 = 2.53 \times 10^{-3} {\rm eV}^2,
\end{eqnarray}
for the neutrino oscillation parameters. Once $d N_{\nu_\mu}/d
E_{\nu_\mu}$ is obtained, we can use the following equation to
calculate the differential muon neutrino flux from the solar DM
annihilation:
\begin{eqnarray}
\frac{d \Phi_{\nu_{\mu}} }{d E_{\nu_{\mu}} } = \frac{\Gamma_{\rm
ANN}}{4 \pi R_{\rm ES}^2} \frac{d N_{\nu_{\mu}} }{d E_{\nu_{\mu}} }
\;, \label{dphide}
\end{eqnarray}
where $R_{\rm ES} = 1.496 \times 10^{13}$ cm is the Earth-Sun
distance. The solar DM annihilation rate $\Gamma_{\rm ANN}$ will be
given in Eq. (\ref{ANN}). In addition, we should also calculate the
differential muon anti-neutrino flux which can be evaluated by an
equation similar to Eq. (\ref{dphide}).

\subsection{Dark matter capture rate and annihilation rate in the Sun}

The halo DM particles can be captured by the Sun via elastic
scattering off solar nuclei. On the other hand, the DM annihilation
in the Sun depletes the DM population. The evolution of the DM
number $N$ in the Sun is given by the following equation
\cite{Griest:1986yu}:
\begin{eqnarray}
\dot{N} = C_\odot - C_E N - C_A N^2 \;, \label{N}
\end{eqnarray}
where the dot denotes differentiation with respect to time. The
solar capture rate $C_\odot$ may be approximately written as
\cite{DM1}
\begin{eqnarray}
C_\odot & \approx & 4.8 \times 10^{24} {\rm s}^{-1} \frac{\rho_{\rm
DM}}{0.3 \, {\rm GeV / cm^3}} \frac{270\, {\rm km/s}}{\bar{v}}
\frac{1 {\rm GeV}}{m_D} \sum_i F_i(m_D) \frac{\sigma_{{{\rm
N}_i}}^{\rm SI}}{10^{-40} {\rm cm^2}} f_i \phi_i S\left(
\frac{m_D}{m_{{\rm N}_i}} \right) \frac{1 {\rm GeV}}{m_{{\rm N}_i}}
\;, \label{capture}
\end{eqnarray}
where $\sigma_{{{\rm N}_i}}^{\rm SI}$ is the spin-independent cross
section of the DM elastic scattering off nucleus ${\rm N}_i$. For
the local DM density $\rho_{\rm DM}$ and the local DM
root-mean-square velocity $\bar{v}$, we take $\rho_{\rm DM} = 0.3 \,
{\rm GeV / cm^3}$ and $\bar{v} = 270 \, {\rm km/s}$. $f_i$ and
$\phi_i$ describe the mass fraction and the distribution of the
element $i$ in the Sun, respectively. $f_i$, $\phi_i$ and the
form-factor suppression $F_i(m_D)$ can be found in Ref. \cite{DM1}.
The function $S(x)$ denotes the kinematic suppression and is given
by
\begin{eqnarray}
S(x) = \left[ \frac{A(x)^{1.5}}{1+A(x)^{1.5}} \right]^{2/3}
\end{eqnarray}
with
\begin{eqnarray}
A(x) = \frac{3x}{2(x-1)^2} \left( \frac{\langle v_{\rm esc}
\rangle}{\bar{v}} \right)^2\;,
\end{eqnarray}
where $\langle v_{\rm esc} \rangle = 1156$ km s$^{-1}$ is a mean
escape velocity. In Eq. (\ref{N}), the term $C_E N$ describes the DM
evaporation rate. For the parameter $C_E$, we adopt the following
approximate formula \cite{Gould:1987ju,Hooper:2008cf}
\begin{eqnarray}
C_E \approx 10^{- 3.5 (m_D/{\rm GeV}) -4} {\rm s}^{-1}
\frac{\sigma_{\rm H}^{\rm SI}}{5 \times 10^{-39} {\rm cm^2}}\;.
\end{eqnarray}
The last term $C_A N^2$ in Eq. (\ref{N}) controls the DM
annihilation rate in the Sun. The coefficient $C_A$ depends on the
thermal-average of the annihilation cross section times the relative
velocity $\langle \sigma v \rangle$ and the DM distribution in the
Sun. To a good approximation,
\begin{eqnarray}
C_A = \frac{\langle \sigma v \rangle}{V_{\rm eff}}\;, \label{sigmav}
\end{eqnarray}
where $V_{\rm eff}$ is the effective volume of the core of the Sun
and is given by \cite{Griest:1986yu}
\begin{eqnarray}
 V_{\rm eff} = 5.8 \times 10^{30} \; {\rm cm^3} \left( \frac{1 {\rm
GeV}}{m_D} \right)^{3/2}.
\end{eqnarray}
It is worthwhile to stress that $\langle \sigma v \rangle$ in Eq.
(\ref{sigmav}) should be evaluated at the solar central temperature
$T_c = 1.4 \times 10^7$ K.

In Refs. \cite{Guo:2010hq} and \cite{Guo:2008si}, we have calculated
the DM-nucleon elastic scattering cross section $\sigma_n^{\rm SI}$
which is equal to $\sigma_{\rm H}^{\rm SI}$. The relation between
$\sigma_{{\rm N}_i}^{\rm SI}$ and $\sigma_{\rm H}^{\rm SI}$  can be
written as
\begin{eqnarray}
\sigma_{{\rm N}_i}^{\rm SI} = A_{{\rm N}_i}^2 \frac{M^2({\rm
N}_i)}{M^2({\rm H})} \sigma_{\rm H}^{\rm SI}\;,
\end{eqnarray}
where $A_{{\rm N}_i}$ is the mass number of the nucleus ${\rm N}_i$
and $M(x) = m_D m_x / (m_D + m_x )$. If $m_D \gg m_{{\rm N}_i}$, we
can easily derive $\sigma_{{\rm N}_i}^{\rm SI} \approx A_{{\rm
N}_i}^4 \, \sigma_{\rm H}^{\rm SI}$. Then one may find that the
solar capture rate by other elements in the Sun is much larger than
that by the hydrogen element although it has the maximal mass
fraction. In terms of relevant formulas  in Refs. \cite{Guo:2010hq}
and \cite{Guo:2008si}, we calculate $\langle \sigma v \rangle$ at
$T_c = 1.4 \times 10^7$ K. Using $\sigma_{\rm H}^{\rm SI}$ and
$\langle \sigma v \rangle$, one can straightly calculate $C_\odot$,
$C_E$ and $C_A$. Then we solve the evolution equation and derive the
solar DM annihilation rate \cite{Griest:1986yu}
\begin{eqnarray}
\Gamma_{\rm ANN} = \frac{1}{2} C_A N^2 = \frac{1}{2} C_\odot \left [
\frac{\tanh (\kappa t_\odot \sqrt{ C_\odot C_A} )}{\kappa + C_E/(2
\sqrt{ C_\odot C_A}) \tanh (\kappa t_\odot \sqrt{ C_\odot C_A} )}
\right]^2 \;, \label{ANN}
\end{eqnarray}
where $\kappa= \sqrt{1+C_E^2/(4 C_\odot C_A)}$ and $t_\odot \simeq
4.5$ Gyr is the age of the solar system. When $C_E$ is small enough
($m_D \gtrsim 4$ GeV), one may neglect the evaporation effect and
obtain
\begin{eqnarray}
\Gamma_{\rm ANN} = \frac{1}{2} C_\odot \tanh^2 (t_\odot \sqrt{
C_\odot C_A} ) \;.
\end{eqnarray}
If $t_\odot \sqrt{ C_\odot C_A} \gg 1$, the DM annihilation rate
reaches equilibrium with the DM capture rate. In this case, we
derive the maximal DM annihilation rate $\Gamma_{\rm ANN} =
C_\odot/2$ which is entirely determined by $C_\odot$. Therefore the
enhanced $\langle \sigma v \rangle$ via the Breit-Wigner resonance
enhancement mechanism \cite{BW} can not affect $\Gamma_{\rm ANN}$.
For $m_D \gtrsim 4$ GeV, we find that most parts of the parameter
space reach or approach the equilibrium except for the resonance
region. It is because that both $\sigma_{n}^{\rm SI}$ and $\langle
\sigma v \rangle$ are very small in this region \cite{Guo:2010hq}.

\subsection{Neutrino induced upgoing muon fluxes in the Super-Kamiokande}

The high energy muon neutrinos from the solar DM annihilation
interact with the Earth rock to produce the upgoing muon flux which
can be detected by the SK detector \cite{SK}. The neutrino induced
muon flux is give by \cite{Gaisser:1984mx}
\begin{eqnarray}
\Phi_\mu  &=&  \int_{E^{\rm SK}_{\rm thr}}^{m_D} d E_{\mu}
\int_{E_{\mu}}^{m_D} d E_{\nu_\mu} \frac{d \Phi_{\nu_{\mu}} }{d
E_{\nu_{\mu}} } \int_0^\infty d L \int_{E_{\mu}}^{E_{\nu_\mu}} d
E_{\mu}' g(L,E_{\mu},E_\mu') \sum_{a = p, n} \frac{d
\sigma_{\nu_\mu}^a (E_{\nu_\mu}, E_\mu')}{d E_\mu'} \rho_a \nonumber
\\ & &  + (\nu_\mu \rightarrow \bar{\nu}_\mu),
\end{eqnarray}
where $\rho_p \approx 1/2 N_A \rho$ and $\rho_n \approx 1/2 N_A
\rho$ are the number densities of protons and neutrons near the
detector, respectively. $N_A$ is the Avogadro's number and $\rho$ is
the density of the rock under the detector. $E_{\rm thr}^{\rm SK} =
1.6$ GeV is the threshold energy of the SK detector. $g(L, E_{\mu},
E_\mu') d E_{\mu}$ is the probability that a muon of initial energy
$E_{\mu}'$ has energy between $E_{\mu}$ and $E_{\mu} + d E_{\mu}$
after propagating a distance $L$ in the rock. For the
charged-current interaction cross sections, we use
\cite{Barger:2007xf}
\begin{eqnarray}
\frac{d \sigma_x^a (E_{x},  E_\mu')}{d E'_\mu} \approx \frac{2 m_p
G_F^2}{\pi} \left(  A_x^a + B_x^a \frac{{E'_\mu}^2}{{E_x}^2} \right)
\;,
\end{eqnarray}
where $A_{\nu_\mu}^{n,p} = 0.25, 0.15$, $B_{\nu_\mu}^{n,p} = 0.06,
0.04$ and $A_{\bar{\nu}_\mu}^{n,p} = B_{\nu_\mu}^{p,n}$,
$B_{\bar{\nu}_\mu}^{n,p} = A_{\nu_\mu}^{p,n}$. The probability $g(L,
E_{\mu}, E_\mu')$ can be obtained from the full Monte Carlo
calculation of muon propagation. Here we use the approximation
formula \cite{Gaisser:1984mx}
\begin{eqnarray}
g(L, E_{\mu}, E_\mu') = \frac{\delta(L - L_0)}{\rho (\alpha+\beta
E_{\mu})}\;,
\end{eqnarray}
with
\begin{eqnarray}
L_0 =  \frac{1}{\rho \beta } \ln \frac{\alpha + \beta E_\mu'
}{\alpha + \beta E_\mu } \;,
\end{eqnarray}
where $\alpha= 2.3 \times 10^{-3} \, {\rm g}^{-1}\, {\rm GeV}\, {\rm
cm}^2 $ and $\beta = 4.4 \times 10^{-6} \, {\rm g}^{-1} \,{\rm
cm}^2$ describe muon energy loss in the standard rock
\cite{Covi:2009xn}. It is shown that this analytic approximation is
good to within $10\%$ or better \cite{Gaisser:1984mx}. Then one can
derive
\begin{eqnarray}
\Phi_\mu = \int_{E^{\rm SK}_{\rm thr}}^{m_D} d E_{\mu} \frac{1}{\rho
(\alpha+\beta E_{\mu})}\int_{E_{\mu}}^{m_D} d E_{\nu_\mu} \frac{d
\Phi_{\nu_{\mu}} }{d E_{\nu_{\mu}} } \int_{E_{\mu}}^{E_{\nu_\mu}} d
E_{\mu}'\sum_{a = p, n} \frac{d \sigma_{\nu_\mu}^a (E_{\nu_\mu},
E_\mu')}{d E_\mu'} \rho_a + (\nu_\mu \rightarrow
\bar{\nu}_\mu)\;.\label{nubar}
\end{eqnarray}
Using a change of variable, we find that the formula in Eq.
(\ref{nubar}) is consistent with that in Ref. \cite{Erkoca:2009by}.

\begin{figure}[htb]
\begin{center}
\includegraphics[scale=0.4]{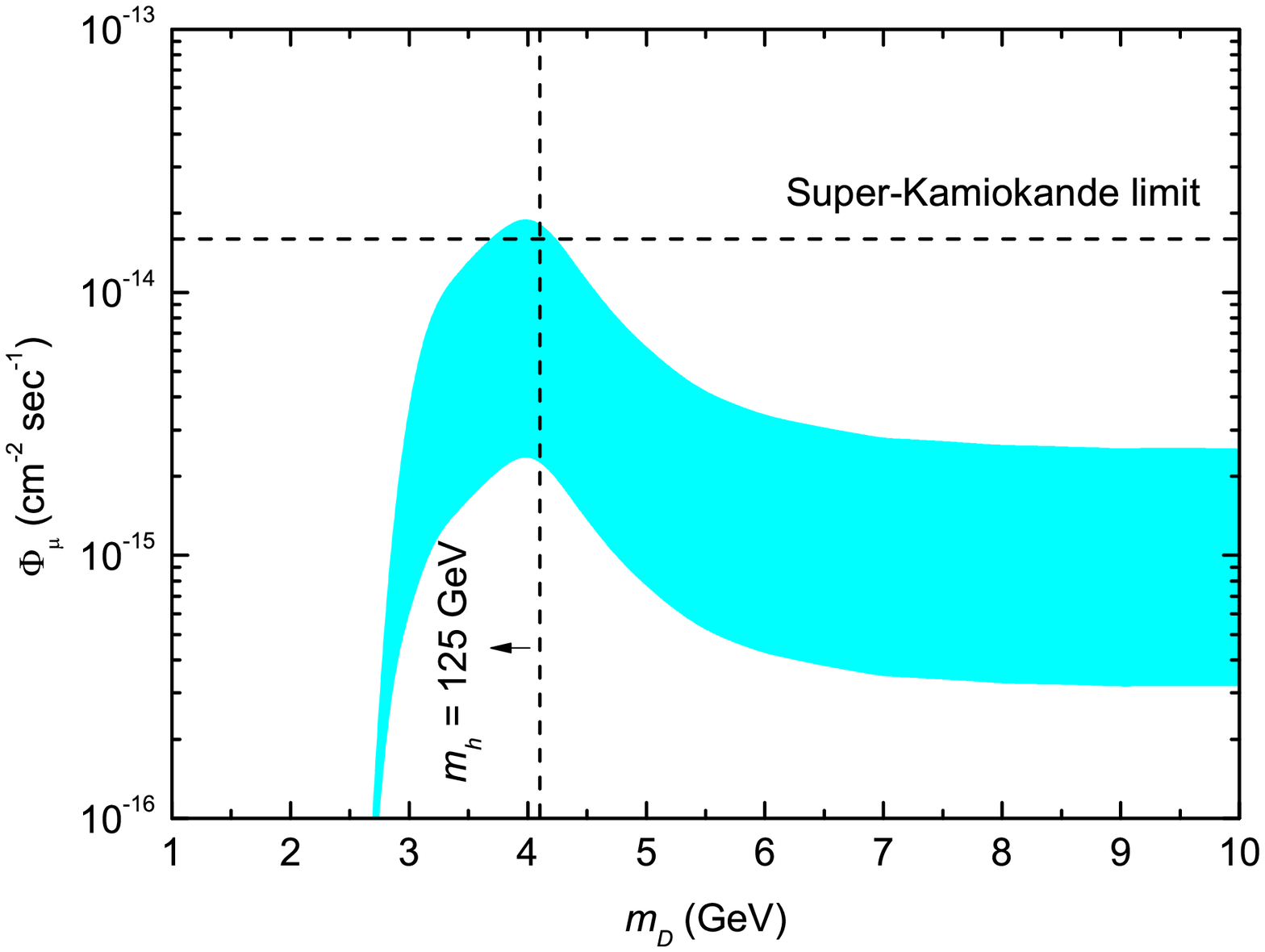}
\includegraphics[scale=0.4]{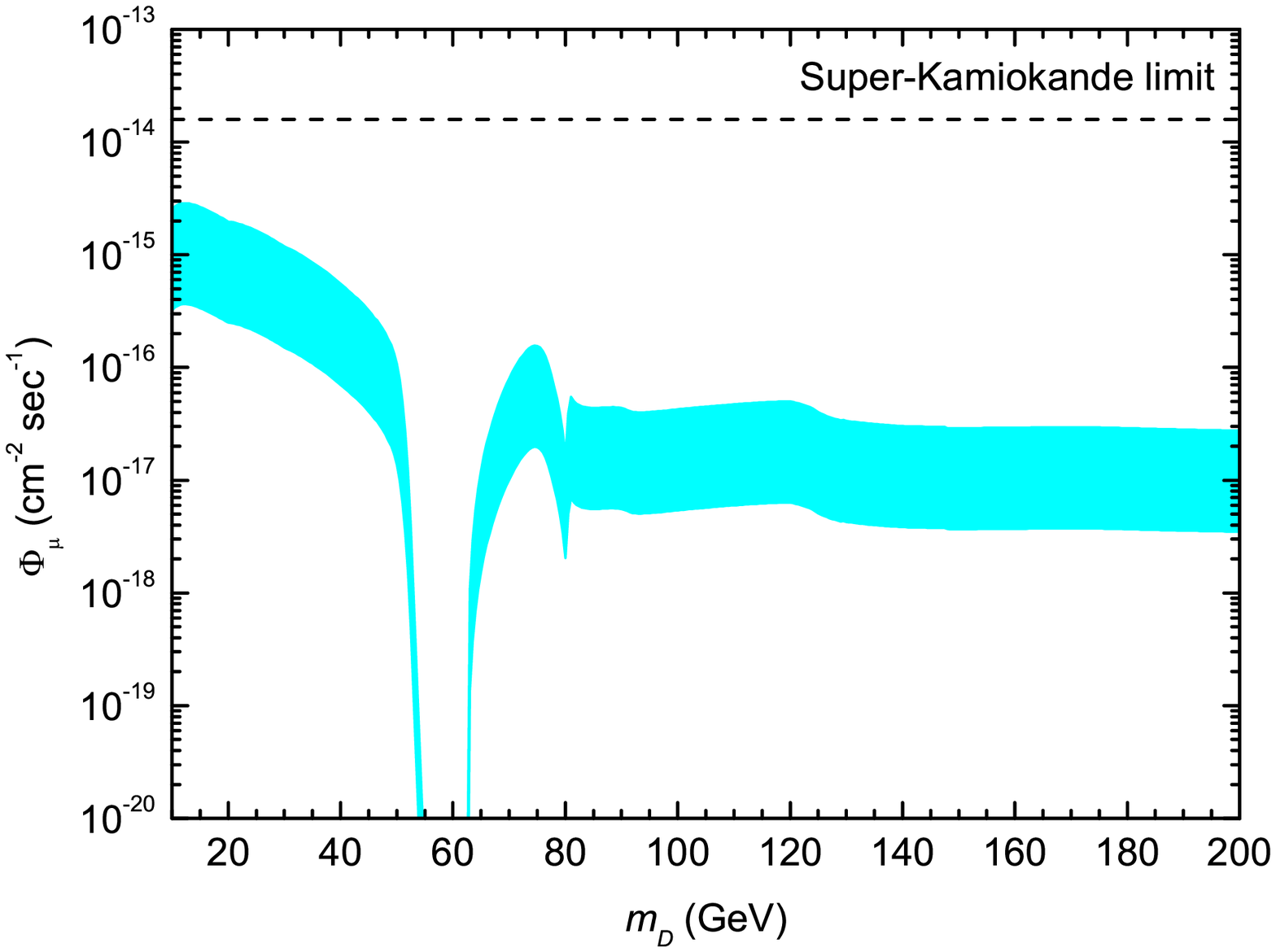}
\includegraphics[scale=0.4]{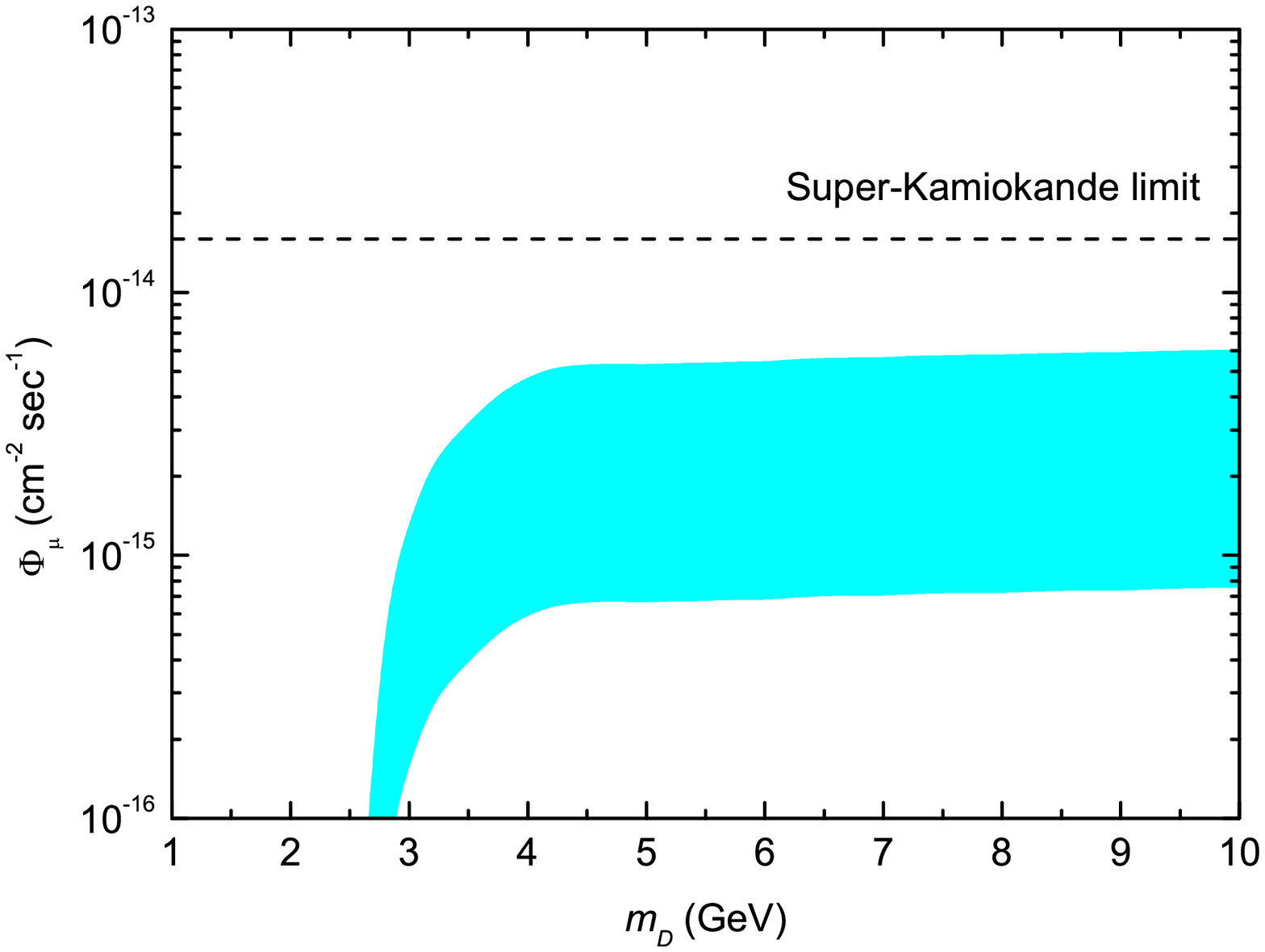}
\includegraphics[scale=0.4]{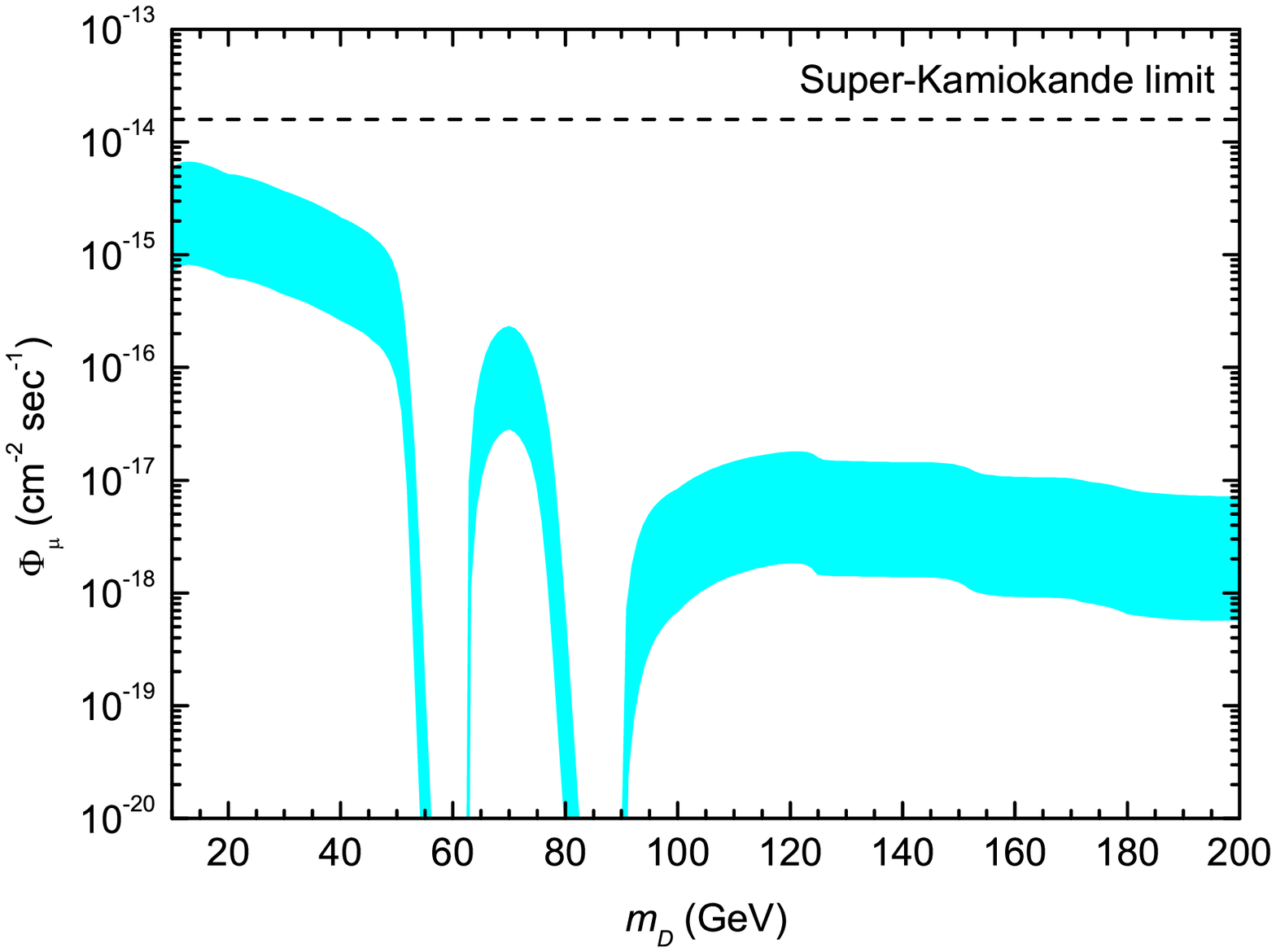}
\end{center}
\caption{ The predicted neutrino induced upgoing muon fluxes for $1
\;{\rm GeV} \leq m_D \leq 200$ GeV in the SSDM-SM (top row) and
SSDM-2HBDM (bottom row). The dashed line denotes the
Super-Kamiokande muon flux limit.} \label{SuperK}
\end{figure}

For the SSDM-SM, we calculate the neutrino induced upgoing muon
fluxes in the Super-Kamiokande with the help of Eqs. (\ref{dphide}),
(\ref{ANN}) and (\ref{nubar}). The numerical results have been shown
in Fig. \ref{SuperK} (top row). Due to the multiple Coulomb
scattering of muons on route to the detector, the final directions
of muons are spread. For $10 \;{\rm GeV} \leq m_D \leq 200$ GeV, the
cone half-angles range from $5^\circ$ to $25^\circ$
\cite{Mori:1993tj}. Therefore we conservatively take $\Phi_\mu \leq
1.6 \times 10^{-14} {\rm cm^{-2}} {\rm s^{-1}}$ (maximal value in
Fig. 8 of Ref. \cite{SK}) for the Super-Kamiokande limit.  It is
clear that our results in the region $3.7 \, {\rm GeV} \leq m_D \leq
4.2$ GeV and $f \gtrsim 0.65$ slightly exceed the Super-Kamiokande
limit. Since the uncertainties in the astrophysics and particle
physics, such as $\rho_{\rm DM}$, $\bar{v}$ and $\alpha$, we can not
claim that the Super-Kamiokande can exclude this region. Notice that
the exceeded region is not consistent with the CDMS (shallow-site
data) results as shown in Fig. \ref{Direct} (top left panel). For
$10 \;{\rm GeV} \leq m_D \leq 200$ GeV, our numerical results in
Fig. \ref{SuperK} (top right panel) show that the predicted muon
fluxes are less than the Super-Kamiokande limit. In Fig. 11 of Ref.
\cite{SK}, the Super-Kamiokande collaboration has also given the
neutrino induced upgoing muon flux limits as a function of the DM
mass. Their simulations assume that $80\%$ of the annihilation
products are from $b \bar{b}$,  $10\%$ from  $c \bar{c}$ and $10\%$
from  $\tau^+ \tau^-$. It is found that $\Phi_\mu \leq 6.4 \times
10^{-15} {\rm cm^{-2}} {\rm s^{-1}}$ at $m_D =200$ GeV \cite{SK}. In
this case, our numerical results are still far less than $6.4 \times
10^{-15} {\rm cm^{-2}} {\rm s^{-1}}$.

For the SSDM-2HBDM, the large Yukawa scale factors $R_l =10$ for
charged leptons can significantly  enhance the branching ratio of
the $\tau^+ \tau^- $ annihilation channel when two DM particles can
not annihilate into $W^+ W^-$ ($m_D < m_W$). Since the produced muon
event numbers from a pair of $\tau^+ \tau^- $ are far larger than
those from $b \bar{b}$ and $c \bar{c}$. Therefore the SSDM-2HBDM
with the enhanced $\tau^+ \tau^- $ branching ratio ($B_{\tau^+
\tau^-} \simeq 53\%$ at $m_D = 10$ GeV) can give larger neutrino
induced upgoing muon fluxes than those in the SSDM-SM even if the
SSDM-2HBDM has the smaller $\sigma_{n}^{\rm SI}$ as shown in Figs.
\ref{Direct} and \ref{SuperK}. If $R_l \gg 10$, one will obtain a
smaller $\lambda_{1, D}$ from the desired DM relic density which
leads to a smaller $\sigma_{n}^{\rm SI}$. In this case, the
SSDM-2HBDM will produce smaller muon fluxes since $R_l \gg 10$ does
not significantly enlarge $B_{\tau^+ \tau^-}$. For $10 \;{\rm GeV}
\leq m_D \leq 200$ GeV, the predicted muon fluxes in the SSDM-2HBDM
are far less than the Super-Kamiokande limit as shown in Fig.
\ref{SuperK} (bottom right panel). Since $\sigma_{n}^{\rm SI}$ in
the SSDM-2HBDM may approach the current experimental upper bound
through adjusting $R_q$, we can roughly evaluate the maximal muon
fluxes from Figs. \ref{Direct} and \ref{SuperK} (bottom right
panels). We find that the maximal neutrino induced upgoing muon
fluxes in the SSDM-2HBDM are still less than the Super-Kamiokande
limit when $m_D
> m_W$.

\subsection{Neutrino induced upgoing muon event rates in the IceCube}

The neutrino induced upgoing muons can also be detected by the
neutrino telescope IceCube \cite{IceCube}. In this subsection, we
use the following formula to calculate the neutrino induced upgoing
muon event rates in the IceCube:
\begin{eqnarray}
N_\mu &= & \int_{E^{\rm IC}_{\rm thr}}^{m_D} d E_{\mu} A_{\rm eff}
(E_\mu) \frac{\langle R(\cos \theta_z) \rangle}{2}
\frac{1}{\rho(\alpha+\beta E_{\mu})}\int_{E_{\mu}}^{m_D} d
E_{\nu_\mu} \frac{d \Phi_{\nu_{\mu}} }{d E_{\nu_{\mu}} }
\int_{E_{\mu}}^{E_{\nu_\mu}} d E_{\mu}' \sum_{a= p, n} \frac{d
\sigma_{\nu_\mu}^a (E_{\nu_\mu},E_\mu')}{d E_\mu'} \rho_a  \nonumber
\\ & & + (\nu_\mu \rightarrow \bar{\nu}_\mu), \label{1/2}
\end{eqnarray}
where $A_{\rm eff}(E_\mu )$ and $E_{\rm thr}^{\rm IC} = 50$ GeV are
the effective area and the threshold energy of the IceCube detector.
To a good approximation, $A_{\rm eff}(E_\mu )$ has a very simple
functional form \cite{GonzalezGarcia:2009jc}
\begin{eqnarray}
A_{\rm eff} (E_\mu  \leq 10^{1.6} {\rm GeV}) & = & 0,   \nonumber \\
A_{\rm eff} ( 10^{1.6} {\rm GeV} <  E_\mu  < 10^{2.8} {\rm GeV}) & = & 0.748 [ \log(E_\mu/{\rm GeV}) - 1.6] \,{\rm km}^2,    \nonumber \\
A_{\rm eff} (E_\mu \geq 10^{2.8} {\rm GeV}) & = & 0.9 + 0.54 [
\log(E_\mu/{\rm GeV}) - 2.8] \,{\rm
 km}^2.
\end{eqnarray}
$R(\cos \theta_z)$ is a phenomenological angular dependence of the
effective area for upgoing muons
\begin{eqnarray}
R(\cos \theta_z) = 0.92 - 0.45 \cos \theta_z  \;,
\end{eqnarray}
where $\theta_z$ is the zenith angle. Considering the change of the
Sun direction, we average $R(\cos \theta_z)$ from $\cos (90^\circ)$
to $\cos (113.43^\circ)$ and derive $\langle R(\cos \theta_z)
\rangle = 1.01$. The factor of $1/2$ in Eq. (\ref{1/2}) accounts for
about $50 \% $ of the time that the Sun is below the horizon. For
the ice, we take $\alpha= 2.7 \times 10^{-3} \, {\rm g}^{-1}\, {\rm
GeV}\, {\rm cm}^2 $, $\beta = 3.3 \times 10^{-6} \, {\rm g}^{-1}
\,{\rm cm}^2$, $\rho_p \approx 5/9 N_A \rho$ and $\rho_n \approx 4/9
N_A \rho$ \cite{Covi:2009xn}.

\begin{figure}[htb]
\begin{center}
\includegraphics[scale=0.4]{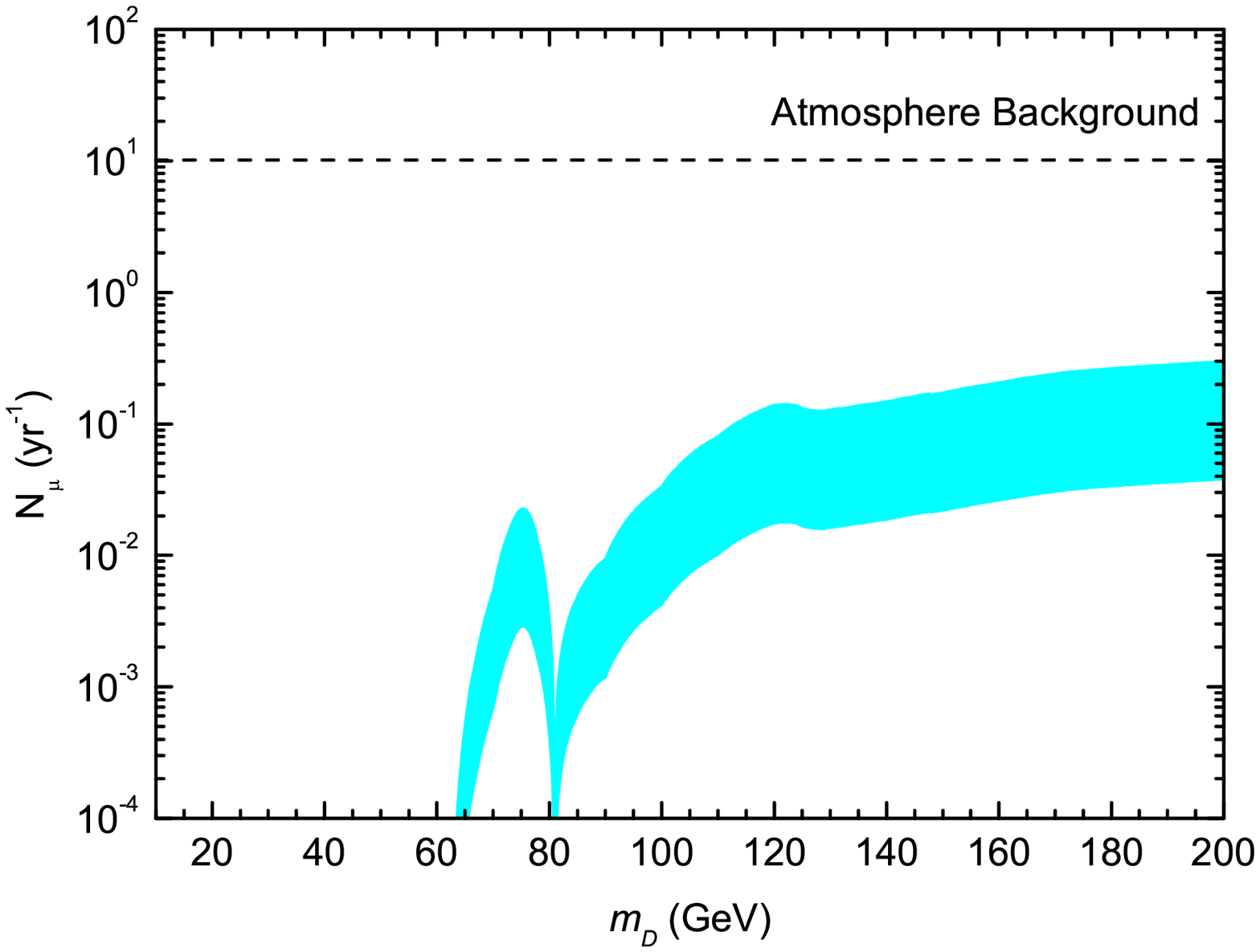}
\includegraphics[scale=0.4]{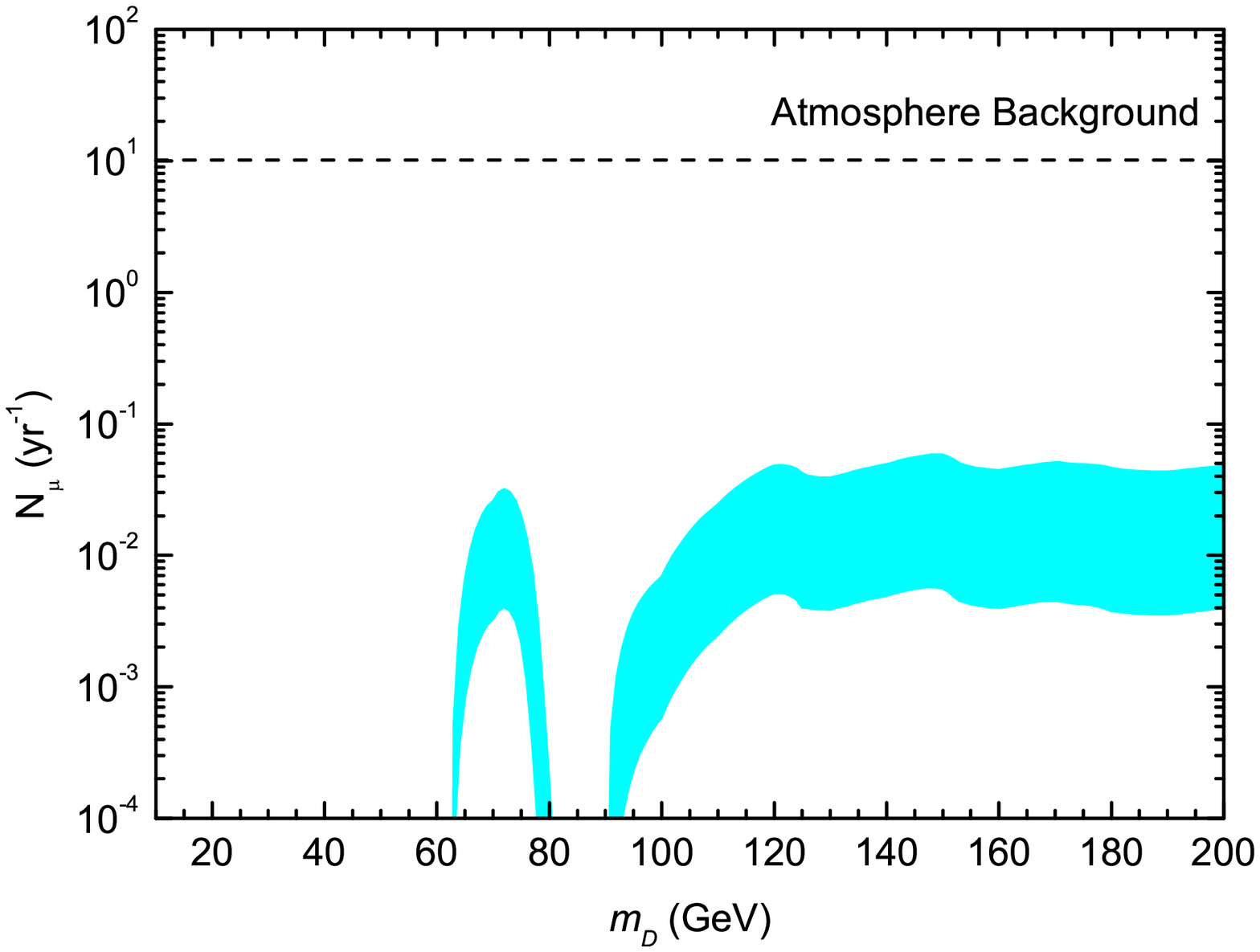}
\end{center}
\caption{ The predicted muon event rates for $10 \;{\rm GeV} \leq
m_D \leq 200$ GeV  in the SSDM-SM (left panel) and SSDM-2HBDM (right
panel). The dashed line denotes the atmosphere background in the
IceCube.} \label{IceCube}
\end{figure}

We use the above formulas to calculate the muon neutrino and muon
anti-neutrino induced upgoing muon event rates as well as the
background from atmosphere neutrinos in the IceCube. The atmosphere
neutrino fluxes $d \Phi_{\nu_{\mu}}/ d E_{\nu_{\mu}} (\cos
\theta_z)$ can be found in Ref. \cite{Honda:2006qj}. For the
atmosphere background, $\langle R(\cos \theta_z) \rangle d
\Phi_{\nu_{\mu}}/ d E_{\nu_{\mu}}$ in Eq. (\ref{1/2}) should be
replaced by $\langle R(\cos \theta_z)  d \Phi_{\nu_{\mu}}/ d
E_{\nu_{\mu}}(\cos \theta_z)\rangle$. In order to reduce the
background from atmosphere neutrinos, we require $E_{\rm thr}^{\rm
IC} \leq E_\mu \leq 200$ GeV and only consider the fluxes observed
along the line of sight to the Sun within the $2^\circ$ half-angle
cone \cite{Liu:2008kz}. Our numerical results have been shown in
Fig. \ref{IceCube}. It is found that the predicted muon event rates
in the SSDM-SM and SSDM-2HBDM are less than the atmosphere
background 10.2 yr$^{-1}$.

\section{Discussions and Conclusions}

In terms of the observed DM abundance,  we can derive the DM-Higgs
couplings $\lambda$ in the SSDM-SM and $\lambda_{1, D}$ in the
SSDM-2HBDM. If $\lambda^2$ and $\lambda_{1, D}^2$ are enlarged by
$X$ times, the spin-independent DM-nucleon elastic scattering cross
section $\sigma_{n}^{\rm SI}$ in the SSDM-SM and SSDM-2HBDM will be
enlarged by the same times as shown in Eqs. (\ref{sigmaSM}) and
(\ref{sigma2HBDM}). Since the DM relic density will be approximately
suppressed by $X$ times, one thus needs to introduce new DM
candidates. In terms of Eq. (\ref{capture}), one may find that the
produced neutrino signals from the DM candidates $S$ and $S_D$ do
not significantly change as the couplings $\lambda$ and $\lambda_{1,
D}$ increase.

In conclusion, we have investigated the singlet scalar dark matter
from direct detections and  high energy neutrino signals via the
solar DM annihilation in the SSDM-SM and SSDM-2HBDM. Firstly, we
consider the uncertainties in the hadronic matrix elements and
recalculate the spin-independent DM-nucleon elastic scattering cross
section $\sigma_{n}^{\rm SI}$. It is found that the current DM
direct detection experiments can exclude the $f \gtrsim 0.63$ region
for $1 \;{\rm GeV} \leq m_D \leq 10$ GeV in the SSDM-SM. The latest
XENON100 may exclude $7 \;{\rm GeV} \lesssim m_D \lesssim 52$ GeV
and a narrow region  $65 \;{\rm GeV} \lesssim m_D \lesssim 80$  GeV
for $m_h$ = 125 GeV even if we take $f = 0.26$. For the SSDM-2HBDM,
we can adjust the Yukawa couplings to avoid the direct detection
limits. Then we numerically calculate the neutrino fluxes from the
DM annihilation in the Sun and the neutrino induced upgoing muon
fluxes in the Super-Kamiokande and IceCube. The predicted muon
fluxes in the region $3.7 \, {\rm GeV} \leq m_D \leq 4.2$ GeV and $f
\gtrsim 0.65$ slightly exceed the Super-Kamiokande limit in the
SSDM-SM. However, this exceeded region can be excluded by the CDMS
(shallow-site data). We find that the SSDM-2HBDM can give larger
muon fluxes than those in the SSDM-SM even if the SSDM-2HBDM has
smaller $\sigma_{n}^{\rm SI}$. This is because that the large Yukawa
scale factors $R_l =10$ for charged leptons can significantly
enhance the branching ratio of the $\tau^+ \tau^- $ annihilation
channel and the produced muon event numbers from a pair of $\tau^+
\tau^- $ are far larger than those from $b \bar{b}$ and $c \bar{c}$.
For the allowed parameter space of the SSDM-SM and SSDM-2HBDM, the
produced muon fluxes in the Super-Kamiokande and muon event rates in
the IceCube are less than the experiment upper bound and atmosphere
background, respectively. The large muon fluxes in $3 \;{\rm GeV}
\lesssim m_D \lesssim 10$ GeV indicate that the future neutrino
experiments can provide constraints on the SSDM-SM and SSDM-2HBDM.

\acknowledgments

This work is supported in part by the National Basic Research
Program of China (973 Program) under Grants No. 2010CB833000; the
National Nature Science Foundation of China (NSFC) under Grants No.
10821504 and No. 10905084; and the Project of Knowledge Innovation
Program (PKIP) of the Chinese Academy of Science.


\begin{thebibliography}{99}



\bibitem{DM1} G. Jungman, M. Kamionkowski and K.
Griest, Phys. Rept. {\bf 267}, 195 (1996).

\bibitem{DM2}
G. Bertone, D. Hooper and J. Silk, Phys. Rept. {\bf 405}, 279
(2005).

\bibitem{WMAP7}
  E.~Komatsu {\it et al.}  [WMAP Collaboration],
  Astrophys.\ J.\ Suppl.\  {\bf 192}, 18 (2011)  [arXiv:1001.4538 [astro-ph.CO]].



\bibitem{Barger:2007xf}
  V.~Barger, W.~Y.~Keung, G.~Shaughnessy and A.~Tregre,
  Phys.\ Rev.\  D {\bf 76}, 095008 (2007)
  [arXiv:0708.1325 [hep-ph]].

\bibitem{Liu:2008kz}
  J.~Liu, P.~f.~Yin and S.~h.~Zhu,
  Phys.\ Rev.\  D {\bf 77}, 115014 (2008)
  [arXiv:0803.2164 [hep-ph]].

\bibitem{Hooper:2008cf}
  D.~Hooper, F.~Petriello, K.~M.~Zurek and M.~Kamionkowski,
  Phys.\ Rev.\  D {\bf 79}, 015010 (2009)
  [arXiv:0808.2464 [hep-ph]].

\bibitem{Erkoca:2009by}
  A.~E.~Erkoca, M.~H.~Reno and I.~Sarcevic,
  Phys.\ Rev.\  D {\bf 80}, 043514 (2009)
  [arXiv:0906.4364 [hep-ph]].

\bibitem{Ellis:2009ka}
  J.~Ellis, K.~A.~Olive, C.~Savage and V.~C.~Spanos,
  Phys.\ Rev.\  D {\bf 81}, 085004 (2010)
  [arXiv:0912.3137 [hep-ph]].

\bibitem{SUN}
  G.~Wikstrom and J.~Edsjo,
  JCAP {\bf 0904}, 009 (2009)
  [arXiv:0903.2986 [astro-ph.CO]];
  V.~Niro, A.~Bottino, N.~Fornengo and S.~Scopel,
  Phys.\ Rev.\  D {\bf 80}, 095019 (2009)
  [arXiv:0909.2348 [hep-ph]];
  J.~Shu, P.~f.~Yin and S.~h.~Zhu,
  Phys.\ Rev.\  D {\bf 81}, 123519 (2010)
  [arXiv:1001.1076 [hep-ph]];
  A.~L.~Fitzpatrick, D.~Hooper and K.~M.~Zurek,
  Phys.\ Rev.\  D {\bf 81}, 115005 (2010)
  [arXiv:1003.0014 [hep-ph]];
  P.~Agrawal, Z.~Chacko, C.~Kilic and R.~K.~Mishra,
  arXiv:1003.5905 [hep-ph];
  V.~Barger, J.~Kumar, D.~Marfatia and E.~M.~Sessolo,
  Phys.\ Rev.\  D {\bf 81}, 115010 (2010)
  [arXiv:1004.4573 [hep-ph]];
  M.~A.~Ajaib, I.~Gogoladze and Q.~Shafi,
  Phys.\ Rev.\ D {\bf 83}, 075017 (2011)  [arXiv:1101.0835 [hep-ph]];  %
  N.~F.~Bell and K.~Petraki,
  JCAP {\bf 1104}, 003 (2011)  [arXiv:1102.2958 [hep-ph]].



\bibitem{SK}
  S.~Desai {\it et al.}  [Super-Kamiokande Collaboration],
  Phys.\ Rev.\  D {\bf 70}, 083523 (2004)
  [Erratum-ibid.\  D {\bf 70}, 109901 (2004)]
  [arXiv:hep-ex/0404025].


\bibitem{IceCube}
  J.~Ahrens {\it et al.}, IceCube Preliminary Design Document
  (2001);
  J.~Ahrens {\it et al.}  [IceCube Collaboration],
  Astropart.\ Phys.\  {\bf 20}, 507 (2004)
  [arXiv:astro-ph/0305196].


\bibitem{SingletDM}
  V.~Silveira and A.~Zee,
  Phys.\ Lett.\  B {\bf 161}, 136 (1985);
  J.~McDonald,
  Phys.\ Rev.\  D {\bf 50}, 3637 (1994)
  [arXiv:hep-ph/0702143];
  X.~G.~He, T.~Li, X.~Q.~Li, J.~Tandean and H.~C.~Tsai,
  Phys.\ Rev.\  D {\bf 79}, 023521 (2009)
  [arXiv:0811.0658 [hep-ph]];
  Phys.\ Lett.\  B {\bf 688}, 332 (2010)
  [arXiv:0912.4722 [hep-ph]];
  S.~Profumo, L.~Ubaldi and C.~Wainwright,
  Phys.\ Rev.\  D {\bf 82}, 123514 (2010)
  [arXiv:1009.5377 [hep-ph]].




\bibitem{Pospelov}
 C.~P.~Burgess, M.~Pospelov and T.~ter Veldhuis,
  Nucl.\ Phys.\  B {\bf 619}, 709 (2001)
  [arXiv:hep-ph/0011335].

\bibitem{Gonderinger:2009jp}
  M.~Gonderinger, Y.~Li, H.~Patel and M.~J.~Ramsey-Musolf,
  JHEP {\bf 1001}, 053 (2010)
  [arXiv:0910.3167 [hep-ph]].

\bibitem{DAMACoGeNT}
  S.~Andreas, T.~Hambye and M.~H.~G.~Tytgat,
  JCAP {\bf 0810}, 034 (2008)
  [arXiv:0808.0255 [hep-ph]];
  A.~Bandyopadhyay, S.~Chakraborty, A.~Ghosal and D.~Majumdar,
  JHEP {\bf 1011}, 065 (2010)  [arXiv:1003.0809 [hep-ph]];
  S.~Andreas, C.~Arina, T.~Hambye, F.~-S.~Ling and M.~H.~G.~Tytgat,
  Phys.\ Rev.\ D {\bf 82}, 043522 (2010)  [arXiv:1003.2595
  [hep-ph]];
  V.~Barger, M.~McCaskey and G.~Shaughnessy,
  Phys.\ Rev.\  D {\bf 82}, 035019 (2010)
  [arXiv:1005.3328 [hep-ph]];
  C.~Arina and M.~H.~G.~Tytgat,
  JCAP {\bf 1101}, 011 (2011)
  [arXiv:1007.2765 [astro-ph.CO]].





\bibitem{Guo:2010hq}
  W.~L.~Guo and Y.~L.~Wu,
  JHEP {\bf 1010}, 083 (2010)
  [arXiv:1006.2518 [hep-ph]], and references therein.

\bibitem{Wu:2007kt}
  Y.~L.~Wu and Y.~F.~Zhou,
  Sci.\ China {\bf G51}, 1808 (2008)
  [arXiv:0709.0042 [hep-ph]];
  Int.\ J.\ Mod.\ Phys.\  A {\bf 23}, 3304 (2008)
  [arXiv:0711.3891 [hep-ph]].


\bibitem{Guo:2010vy}
  W.~L.~Guo, Y.~L.~Wu and Y.~F.~Zhou,
  Phys.\ Rev.\  D {\bf 81}, 075014 (2010)
  [arXiv:1001.0307 [hep-ph]].



\bibitem{Guo:2008si}
  W.~L.~Guo, Y.~L.~Wu and Y.~F.~Zhou,
  Phys.\ Rev.\  D {\bf 82}, 095004 (2010)
  [arXiv:1008.4479 [hep-ph]];
  W.~L.~Guo, L.~M.~Wang, Y.~L.~Wu, Y.~F.~Zhou and C.~Zhuang,
  Phys.\ Rev.\  D {\bf 79}, 055015 (2009)
  [arXiv:0811.2556 [hep-ph]].

\bibitem{LHC}
  ATLAS Collaboration, ATLAS-CONF-2012-093; CMS Collaboration,
  CMS-PAS-HIG-12-020.

\bibitem{miss}
We miss a factor $1/2$ for $a_q$ in Ref. \cite{Guo:2010hq}.
Therefore the predicted direct detection cross sections in Ref.
\cite{Guo:2010hq} correspond to the $f =0.7$ case in this paper.


\bibitem{Giedt:2009mr}
  J.~Giedt, A.~W.~Thomas and R.~D.~Young,
  Phys.\ Rev.\ Lett.\  {\bf 103}, 201802 (2009)
  [arXiv:0907.4177 [hep-ph]].


\bibitem{DAMA}
  R.~Bernabei {\it et al.}  [DAMA Collaboration],
  Eur.\ Phys.\ J.\  C {\bf 56}, 333 (2008)
  [arXiv:0804.2741 [astro-ph]].


\bibitem{CoGeNT}
  C.~E.~Aalseth {\it et al.}  [CoGeNT Collaboration],
  Phys.\ Rev.\ Lett.\  {\bf 106}, 131301 (2011)  [arXiv:1002.4703 [astro-ph.CO]].




\bibitem{Hooper:2010uy}
  D.~Hooper, J.~I.~Collar, J.~Hall and D.~McKinsey,
  Phys.\ Rev.\  D {\bf 82}, 123509 (2010)
  [arXiv:1007.1005 [hep-ph]].


\bibitem{CDMSII}
  Z.~Ahmed {\it et al.}  [CDMS-II Collaboration],
  Phys.\ Rev.\ Lett.\  {\bf 106}, 131302 (2011)  [arXiv:1011.2482
  [astro-ph.CO]];
  Science {\bf 327}, 1619 (2010)
  [arXiv:0912.3592 [astro-ph.CO]].



\bibitem{CDMS}
  D.~S.~Akerib {\it et al.}  [CDMS Collaboration],
  Phys.\ Rev.\  D {\bf 82}, 122004 (2010)
  [arXiv:1010.4290 [astro-ph.CO]].


\bibitem{CRESST}
  G.~Angloher {\it et al.},
  Astropart.\ Phys.\  {\bf 18}, 43 (2002).




\bibitem{TEXONO}
  S.~T.~Lin {\it et al.}  [TEXONO Collaboration],
  Phys.\ Rev.\  D {\bf 79}, 061101 (2009)
  [arXiv:0712.1645 [hep-ex]].



\bibitem{XENON100}
  E.~Aprile {\it et al.}  [XENON100 Collaboration],
  arXiv:1207.5988 [astro-ph.CO].



\bibitem{CDMS100}
J. Cooley, SLAC seminar on Dec. 17, 2009; L. Hsu, Fermilab seminar
on Dec. 17, 2009.


\bibitem{XENON1T} Elena Aprile, XENON1T: a ton scale Dark Matter Experiment ,
presented at UCLA Dark Matter 2010, February 26, 2010. The XENON1000
project in China has been supported in part by the National Basic
Research Program of China (973 Program).

\bibitem{Pospelov:1996fq}
  M.~E.~Pospelov,
  Phys.\ Rev.\  D {\bf 56}, 259 (1997)
  [arXiv:hep-ph/9611422];
  Y.~Zhang, H.~An, X.~Ji and R.~N.~Mohapatra,
  Phys.\ Rev.\  D {\bf 76}, 091301 (2007)
  [arXiv:0704.1662 [hep-ph]];
  A.~Maiezza, M.~Nemevsek, F.~Nesti and G.~Senjanovic,
  Phys.\ Rev.\  D {\bf 82}, 055022 (2010)
  [arXiv:1005.5160 [hep-ph]].


\bibitem{WimpSim}
J. Edsjo, WimpSim Neutrino Monte Carlo,
http://www.physto.se/~edsjo/wimpsim/;
  M.~Blennow, J.~Edsjo and T.~Ohlsson,
  JCAP {\bf 0801}, 021 (2008)
  [arXiv:0709.3898 [hep-ph]].

\bibitem{Pythia}
  T.~Sjostrand, S.~Mrenna and P.~Z.~Skands,
  JHEP {\bf 0605}, 026 (2006)
  [arXiv:hep-ph/0603175].

\bibitem{Nusigma}
J. Edsjo, Nusigma 1.16.


\bibitem{DarkSUSY}
  P.~Gondolo, J.~Edsjo, P.~Ullio, L.~Bergstrom, M.~Schelke and E.~A.~Baltz,
  JCAP {\bf 0407}, 008 (2004)
  [arXiv:astro-ph/0406204].

\bibitem{Bahcall:2004pz}
  J.~N.~Bahcall, A.~M.~Serenelli and S.~Basu,
  Astrophys.\ J.\  {\bf 621}, L85 (2005)
  [arXiv:astro-ph/0412440].


\bibitem{An:2012eh}
  F.~P.~An {\it et al.}  [DAYA-BAY Collaboration],
  Phys.\ Rev.\ Lett.\  {\bf 108}, 171803 (2012)  [arXiv:1203.1669 [hep-ex]].

\bibitem{Tortola:2012te}
  D.~V.~Forero, M.~Tortola and J.~W.~F.~Valle,
  arXiv:1205.4018 [hep-ph].


\bibitem{Griest:1986yu}
  K.~Griest and D.~Seckel,
  Nucl.\ Phys.\  B {\bf 283}, 681 (1987)
  [Erratum-ibid.\  B {\bf 296}, 1034 (1988)].


\bibitem{Gould:1987ju}
  A.~Gould,
  Astrophys.\ J.\  {\bf 321}, 560 (1987).

\bibitem{BW}
  D.~Feldman, Z.~Liu and P.~Nath,
  Phys.\ Rev.\  D {\bf 79}, 063509 (2009)
  [arXiv:0810.5762 [hep-ph]];
  M.~Ibe, H.~Murayama and T.~T.~Yanagida,
  Phys.\ Rev.\  D {\bf 79}, 095009 (2009)
  [arXiv:0812.0072 [hep-ph]];
  W.~L.~Guo and Y.~L.~Wu,
  Phys.\ Rev.\  D {\bf 79}, 055012 (2009)
  [arXiv:0901.1450 [hep-ph]].

\bibitem{Gaisser:1984mx}
  T.~K.~Gaisser and T.~Stanev,
  Phys.\ Rev.\  D {\bf 30}, 985 (1984).


\bibitem{Covi:2009xn}
  L.~Covi, M.~Grefe, A.~Ibarra and D.~Tran,
  JCAP {\bf 1004}, 017 (2010)
  [arXiv:0912.3521 [hep-ph]].


\bibitem{Mori:1993tj}
  M.~Mori {\it et al.}  [KAMIOKANDE Collaboration],
  Phys.\ Rev.\  D {\bf 48}, 5505 (1993).


\bibitem{GonzalezGarcia:2009jc}
  M.~C.~Gonzalez-Garcia, F.~Halzen and S.~Mohapatra,
  Astropart.\ Phys.\  {\bf 31}, 437 (2009)
  [arXiv:0902.1176 [astro-ph.HE]].

\bibitem{Honda:2006qj}
  M.~Honda, T.~Kajita, K.~Kasahara, S.~Midorikawa and T.~Sanuki,
  Phys.\ Rev.\  D {\bf 75}, 043006 (2007)
  [arXiv:astro-ph/0611418].



\end{thebibliography}
\end{document}